%% file: bare_jrnl.tex
\documentclass[final,journal]{IEEEtran}

\usepackage{cite}
\usepackage{graphicx}
\usepackage{subcaption}
\usepackage{lipsum}
\usepackage{epsfig}
\usepackage{epstopdf} 
\usepackage{multirow}
\usepackage{pbox}
\usepackage{booktabs}

\graphicspath{{./figures/}}
\usepackage{xcolor}
\usepackage{color,soul}
\usepackage[official]{eurosym}
\usepackage[nolist]{acronym}
\usepackage{pifont}%
\ifCLASSINFOpdf
\else
\fi
\usepackage{todonotes}
\usepackage{multirow}
\usepackage{booktabs}
\usepackage{longtable}
\usepackage{supertabular}

\begin{document}
%
\title{An Overview on Application of Machine Learning Techniques in Optical Networks}
%
%
%

\author{Francesco Musumeci,~\IEEEmembership{Member,~IEEE,}
        Cristina Rottondi,~\IEEEmembership{Member,~IEEE,}
        Avishek Nag,~\IEEEmembership{Member,~IEEE,}
        Irene Macaluso,~
        Darko Zibar,~\IEEEmembership{Member,~IEEE,}
        Marco Ruffini,~\IEEEmembership{Senior~Member,~IEEE,}
        and~Massimo Tornatore,~\IEEEmembership{Senior~Member,~IEEE}
\thanks{Francesco Musumeci and Massimo Tornatore are with Politecnico di Milano, Italy, e-mail: francesco.musumeci@polimi.it, massimo.tornatore@polimi.it}
\thanks{Cristina Rottondi is with Dalle Molle Institute for Artificial Intelligence, Switzerland, email: cristina.rottondi@supsi.ch.}
\thanks{Avishek Nag is with University College Dublin, Ireland, email: avishek.nag@ucd.ie.}
\thanks{Irene Macaluso and Marco Ruffini are with Trinity College Dublin, Ireland, email: macalusi@tcd.ie, ruffinm@tcd.ie.}
\thanks{Darko Zibar is with Technical University of Denmark, Denmark, email: dazi@fotonik.dtu.dk.}
}

\maketitle

\begin{abstract}

Today's telecommunication networks have become sources of enormous amounts of widely heterogeneous data.
This information can be retrieved from network traffic traces, network alarms, signal quality indicators, users' behavioral data, etc. Advanced mathematical tools are required to extract meaningful information from these data and take decisions pertaining to the proper functioning of the networks from the network-generated data. Among these mathematical tools, Machine Learning (ML) is regarded as one of the most promising methodological approaches to perform network-data analysis and enable automated network self-configuration and fault management.

The adoption of ML techniques in the field of optical communication networks is motivated by the unprecedented growth of network complexity faced by optical networks in the last few years. Such complexity increase is due to the introduction of a huge number of adjustable and interdependent system parameters (e.g., routing configurations, modulation format, symbol rate, coding schemes, etc.) that are enabled by the usage of coherent transmission/reception technologies, advanced digital signal processing and compensation of nonlinear effects in optical fiber propagation. 

In this paper we provide an overview of the application of ML to optical communications and networking. We classify and survey relevant literature dealing with the topic, and we also provide an introductory tutorial on ML for researchers and practitioners interested in this field.
Although a good number of research papers have recently appeared, the application of ML to optical networks is still in its infancy: to stimulate further work in this area, we conclude the paper proposing new possible research directions.

\end{abstract}


%
\IEEEpeerreviewmaketitle

\input{sections/introduction}

\input{sections/Overview_of_ML_methods}
\input{sections/Optical_Networks_and_Systems_Problems_and_their_Mapping_into_ML_problems}

\input{sections/Detailed_survey_of_ML_in_Physical_Layer_Problems}
\input{sections/Detailed_survey_of_ML_in_Networking_Problems}
\input{sections/Comparison}
\input{sections/Discussion_and_Future_Directions}

\input{sections/Conclusion}
\input{sections/glossary}
\bibliographystyle{IEEEtran}
\bibliography{bibtex/bib/bibliography}




\end{document}

%% file: sections/introduction.tex
\section{Introduction}
\label{intro}

Machine learning (ML) is a branch of Artificial Intelligence that pushes forward the idea that, by giving access to the right data, machines can learn by themselves how to solve a specific problem \cite{marsland2015machine}. By leveraging complex mathematical and statistical tools, ML renders machines capable of performing independently intellectual tasks that have been traditionally solved by human beings. This idea of automating complex tasks has generated high interest in the networking field, on the expectation that several activities involved in the design and operation of communication networks can be offloaded to machines. Some applications of ML in different networking areas have already matched these expectations in areas such as intrusion detection \cite{buczak2016survey}, traffic classification \cite{nguyen2008survey}, cognitive radios \cite{bkassiny2013survey}.

Among various networking areas, in this paper we focus on ML for optical networking. Optical networks constitute the basic physical infrastructure of all large-provider networks worldwide, thanks to their high capacity, low cost and many other attractive properties \cite{mukherjee2006optical}. They are now penetrating new important telecom markets as datacom \cite{decusatis2014optical} and the access segment \cite{song2010long}, and there is no sign that a substitute technology might appear in the foreseeable future. Different approaches to improve the performance of optical networks have been investigated, such as routing, wavelength assignment, traffic grooming and survivability \cite{974667, 751461}.

In this paper we give an overview of the application of ML to optical networking. Specifically, the contribution of the paper is twofold, namely, $i$) we provide an introductory tutorial on the use of ML methods and on their application in the optical networks field, and $ii$) we survey the existing work dealing with the topic, also performing a classification of the various use cases addressed in literature so far.
We cover both the areas of optical communication and optical networking to potentially stimulate new cross-layer research directions. In fact, ML application can be  useful especially in cross-layer settings, where data analysis at physical layer, e.g., monitoring Bit Error Rate (BER), can trigger changes at network layer, e.g., in routing, spectrum and modulation format assignments. The application of ML to optical communication and networking is still in its infancy and the literature survey included in this paper aims at providing an introductory reference for researchers and practitioners willing to get acquainted with existing ML applications as well as to investigate new research directions. 

A legitimate question that arises in the optical networking field today is: why machine learning, a methodological area that has been applied and investigated for at least three decades, is only gaining momentum now? The answer is certainly very articulated, and it most likely involves not purely technical aspects \cite{light_reading}. From a technical perspective though, recent technical progress at both optical communication system and network level is at the basis of an unprecedented growth in the complexity of optical networks. 

On a system side, while optical channel modeling has always been complex, the recent adoption of coherent technologies \cite{huawei} has made modeling even more difficult by introducing a plethora of adjustable design parameters (as modulation formats, symbol rates, adaptive coding rates and flexible channel spacing) to optimize transmission systems in terms of bit-rate transmission distance product. In addition, what makes this optimization even more challenging is that the optical channel is highly nonlinear.  

From a networking perspective, the increased complexity of the underlying transmission systems is reflected in a series of advancements in both data plane and control plane. 
At data plane, the Elastic Optical Network (EON) concept \cite{gerstel2012elastic,chatterjee2015routing,talebi2014spectrum,zhang2013survey} has emerged as a novel optical network architecture able to respond to the increased need of elasticity in allocating optical network resources. In contrast to traditional fixed-grid Wavelength Division Multiplexing (WDM) networks, EON offers flexible (almost continuous) bandwidth allocation. Resource allocation in EON can be performed to adapt to the several above-mentioned decision variables made available by new transmission systems, including different transmission techniques, such as Orthogonal Frequency Division Multiplexing (OFDM), Nyquist WDM (NWDM), transponder types (e.g., BVT\footnote{For a complete list of acronyms, the reader is referred to the Glossary at the end of the paper.}, S-BVT), modulation formats (e.g., QPSK, QAM), and coding rates. This flexibility makes the resource allocation problems much more challenging for network engineers. 
At control plane, dynamic control, as in Software-defined networking (SDN), promises to enable long-awaited on-demand reconfiguration and virtualization. Moreover, reconfiguring the optical substrate poses several challenges in terms of, e.g., network re-optimization, spectrum fragmentation, amplifier power settings, unexpected penalties due to non-linearities, which call for strict integration between the control elements (SDN controllers, network orchestrators) and optical performance monitors working at the equipment level. 

All these “degrees of freedom” and limitations do pose severe challenges to system and network engineers when it comes to deciding what the best system and/or network design is.
Machine learning is currently perceived as a paradigm shift for the design of future optical networks and systems. These techniques should allow to infer, from data obtained by various types of monitors (e.g., signal quality, traffic samples, etc.), useful characteristics that could not be easily or directly measured. Some envisioned applications in the optical domain include fault prediction, intrusion detection, physical-flow security, impairment-aware routing, low-margin design, traffic-aware capacity reconfigurations, but many others can be envisioned and will be surveyed in the next sections.

The survey is organized as follows. In Section \ref{overview}, we overview some preliminary ML concepts, focusing especially on those targeted in the following sections. In Section \ref{sec:3} we discuss the main motivations behind the application of ML in the optical domain and we classify the main areas of applications. In Section \ref{sec:4} and Section \ref{sec:5}, we classify and summarize a large number of studies describing applications of ML at the transmission layer and network layer. In Section \ref{sec:comparison}, we quantitatively overview a selection of existing papers, identifying, for some of the applications described in Section \ref{sec:3}, the ML algorithms which demonstrated higher effectiveness for each specific use case, and the performance metrics considered for the algorithms evaluation.
Finally, Section \ref{sec:6} discusses some possible open areas of research and future directions, whereas Section \ref{sec:conc} concludes the paper.

%% file: sections/Overview_of_ML_methods.tex
\section{Overview of machine learning methods used in optical networks}

\label{overview}
\begin{figure}[t!]
\centering
  \subcaptionbox{Supervised Learning: the algorithm is trained on  dataset that consists of paths, wavelengths, modulation and the corresponding BER. Then it extrapolates the BER in correspondence to new inputs.}[.98\linewidth][c]{%
    \includegraphics[width=.98\linewidth]{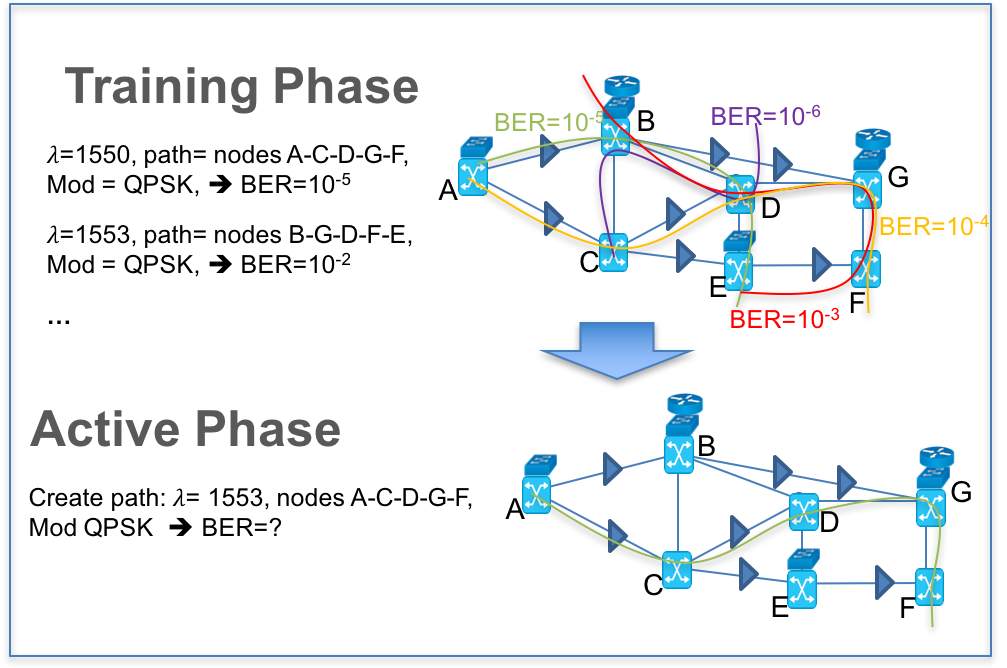}}\quad
  \subcaptionbox{Unsupervised Learning: the algorithm identifies unusual patterns  in the data, consisting of wavelengths, paths, BER, and modulation.}
  [.98\linewidth][c]{%
    \includegraphics[width=.98\linewidth]{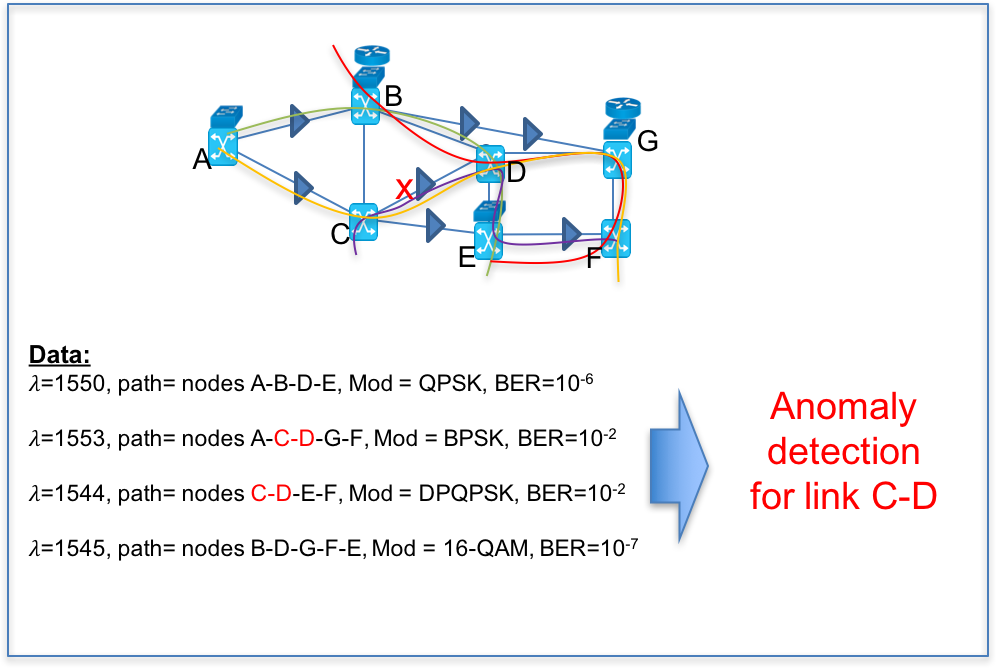}}\quad
    \subcaptionbox{Reinforcement Learning: the algorithm learns by receiving feedback on the effect of modifying some parameters, e.g. the power and the modulation}
    [0.95\linewidth][c]{%
\includegraphics[width=.98\linewidth]{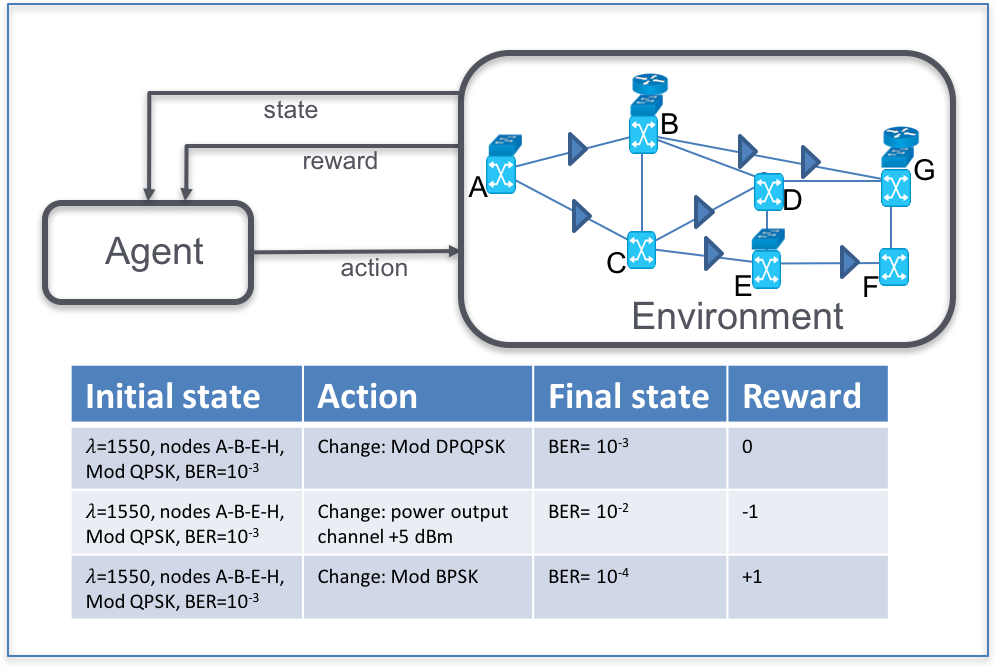}}
\caption{Overview of machine learning algorithms applied to optical networks.}
  \label{fig:ML_methods}
\end{figure}

This section provides an overview of some of the most popular algorithms that are commonly classified as machine learning. The literature on ML is so extensive that even a superficial overview of all the main ML approaches goes far beyond the possibilities of this section, and the readers can refer to a number of fundamental books on the subjects \cite{bishop2006pattern,duda2012pattern,friedman2001elements,sutton1998reinforcement,murphy2014}. However, in this section we provide a high level view of the main ML techniques that are used in the work we reference in the remainder of this paper. We here provide the reader with some basic insights that might help better understand the remaining parts of this survey paper.
We divide the algorithms in three main categories, described in the next sections, which are also represented in Fig. \ref{fig:ML_methods}: supervised learning, unsupervised learning and reinforcement learning. Semi-supervised learning, a hybrid of supervised and unsupervised learning, is also introduced.
ML algorithms have been successfully applied to a wide variety of problems. Before delving into  the different ML methods, it is worth pointing out that, in the context of telecommunication networks, there has been over a decade of research on the application of ML techniques to wireless networks, ranging from opportunistic spectrum access \cite{macaluso2013complexity} to channel estimation and signal detection in OFDM systems \cite{ye2017power}, to Multiple-Input-Multiple-Output communications \cite{o2017deep}, and dynamic frequency reuse \cite{icc2018}.

\subsection{Supervised learning}

Supervised learning is used in a variety of applications, such as speech recognition, spam detection and object recognition.
The goal is to predict the value of one or more output variables given the value of a vector of input variables $\mathbf{x}$. The output variable can be a continuous variable (regression problem) or a discrete variable (classification problem). A training data set comprises $N$ samples of the input variables and the corresponding output values. Different learning methods construct a function $y(\mathbf{x})$ that allows to predict the value of the output variables in correspondence to a new value of the inputs. 
Supervised learning can be broken down into two main classes, described below: \textit{parametric models}, where the number of parameters to use in the model is fixed, and \textit{non-parametric models}, where their number is dependent on the training set.

\subsubsection{Parametric models}
In this case, the function $y$ is a combination of a fixed number of parametric basis functions. These models use training data to estimate a fixed set of parameters $\mathbf{w}$. After the learning stage, the training data can be discarded since the prediction in correspondence to new inputs is computed using only the learned  parameters $\mathbf{w}$. Linear models for regression and classification, which consist of a linear combination of fixed nonlinear basis functions, are the simplest parametric models in terms of analytical and computational properties. Many different choices are available for the basis functions: from polynomial to Gaussian, to sigmoidal, to Fourier basis, etc. In case of multiple output values,  it is possible to use separate basis functions for each component of the output or, more commonly, apply the same set of basis functions for all the components. Note that these models are linear in the parameters  $\mathbf{w}$, and this linearity results in a number of advantageous properties, e.g., closed-form solutions to the least-squares problem.  However, their applicability is limited to problems with low-dimensional input space.
In the remainder of this subsection we focus on neural networks (NNs)\footnote{Note that NNs are often referred to as Artificial Neural Networks (ANNs). In this paper we use these two terms interchangeably.}, since they are the most successful example of parametric models.


\begin{figure}
\centering
    \includegraphics[width=9cm]{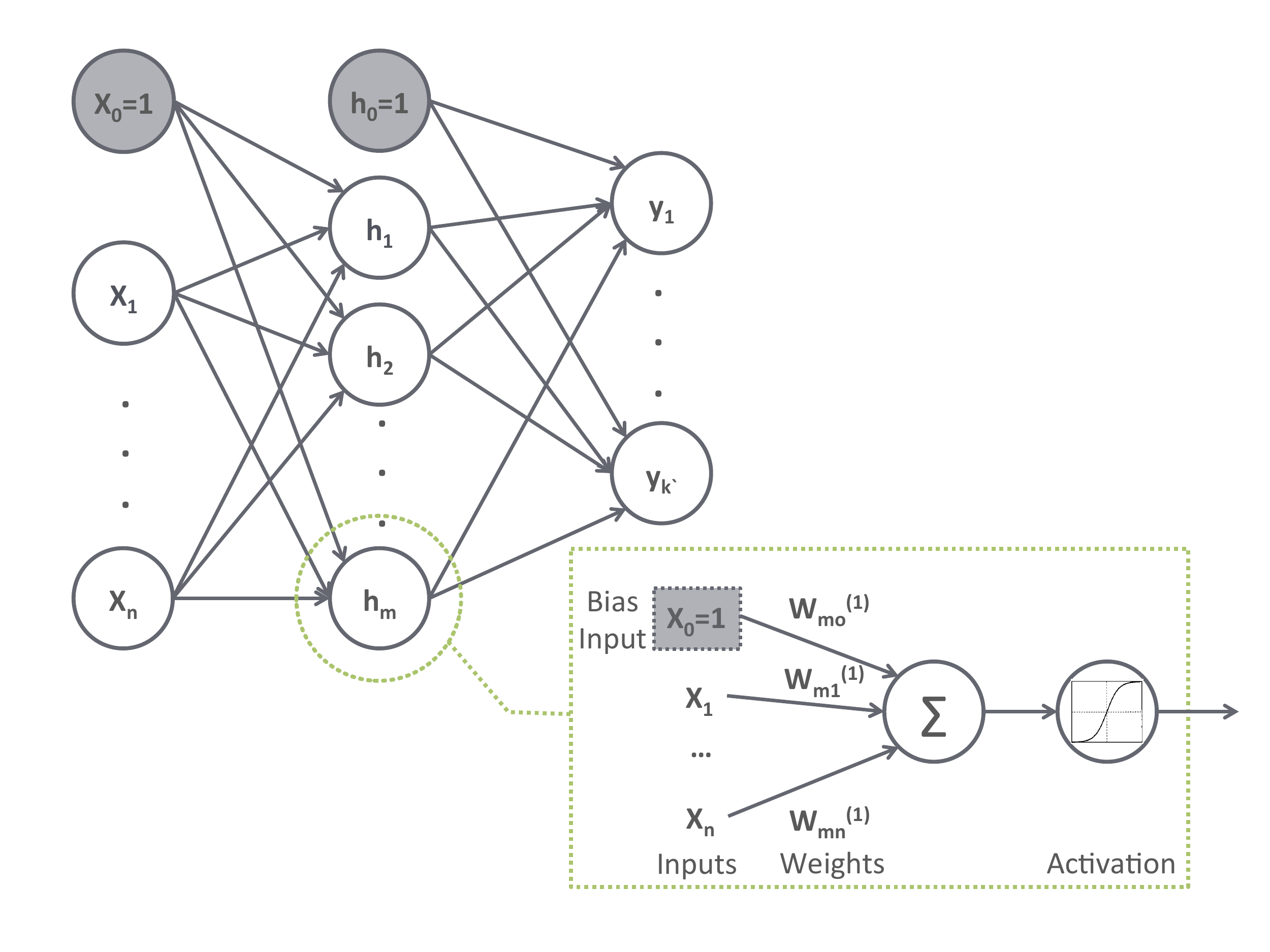}
    \caption{Example of a NN with two layers of adaptive parameters. The bias parameters of the input layer and the hidden layer are represented as weights from additional units with fixed value $1$ ($x_0$ and $h_0$). } \label{fig:nn}
\end{figure}

NNs apply a series of functional transformations to the inputs (see chapter V in \cite{bishop2006pattern}, chapter VI in \cite{duda2012pattern}, and chapter XVI in \cite{murphy2014}). A NN is a network of \textit{units} or \textit{neurons}. The basis function or activation function used by each unit is a nonlinear function of a linear combination of the unit's inputs. Each neuron has a bias parameter that allows for any fixed offset in the data. The bias is incorporated in the set of parameters by adding a dummy input of unitary value to each unit (see Figure \ref{fig:nn}). The coefficients of the linear combination are the parameters $\mathbf{w}$ estimated during the training.  The most commonly used nonlinear functions are the logistic sigmoid and the hyperbolic tangent. The activation function of the output units of the NN is the identity function, the logistic sigmoid function, and the softmax function, for regression, binary classification, and multiclass classification  problems respectively.
 
Different types of connections between the units result in different NNs with distinct characteristics.  All units between the inputs and output of the NN are called \textit{hidden units}. In the case of a NN, the network is a directed acyclic graph. Typically, NNs are organized in layers, with units in each layer receiving inputs only from units in the immediately preceding layer and forwarding their output only to the immediately following layer. NNs with one layer of hidden units and linear output units can approximate arbitrary well any continuous function on a compact domain provided that a sufficient number of hidden units is used \cite{hornik1991approximation}.

Given a training set, a NN is trained by minimizing  an error function with respect to the set of parameters $\mathbf{w}$. Depending on the type of problem and the corresponding choice of activation function of the output units, different error functions are used. Typically in case of regression models, the sum of square error is used, whereas for  classification the cross-entropy error function is adopted. It is important to note that the error function is a non convex function of the network parameters, for which multiple optimal local solutions exist. Iterative numerical methods based on gradient information are the most common methods used to find the vector $\mathbf{w}$ that minimizes the error function. For a NN the error backpropagation algorithm, which provides an efficient method for evaluating the derivatives of the error function with respect to $\mathbf{w}$, is the most commonly used. 

We should at this point mention that, before training the network, the training set is typically pre-processed by applying a linear transformation to rescale each of the input variables independently in case of continuous data or discrete ordinal data. The transformed variables have zero mean and unit standard deviation. The same procedure is applied to the target values in case of regression problems. In case of discrete categorical data, a 1-of-K coding scheme is used. This form of pre-processing is known as feature normalization and it is used before training most ML algorithms since most models are designed with the assumption that all features have comparable scales\footnote{However, decision tree based models are a well-known exception.}.

\subsubsection{Nonparametric models}
In nonparametric methods the number of parameters depends on the training set. These methods keep a subset or the entirety of the training data and use them during prediction. The most used approaches are k-nearest neighbor models (see chapter IV in \cite{duda2012pattern}) and support vector machines (SVMs) (see chapter VII in \cite{bishop2006pattern} and chapter XIV in \cite{murphy2014}). Both can be used for regression and classification problems.

In the case of k-nearest neighbor methods, all training data samples are stored (training phase). During prediction, the k-nearest samples to the new input value are retrieved. For classification problem, a voting mechanism is used; for regression problems, the mean or median of the k nearest samples provides the prediction. To select the best value of k, cross-validation \cite{hastie2009} can be used. 
Depending on the dimension of the training set, iterating through all samples to compute the closest k neighbors might not be feasible. In this case, k-d trees or locality-sensitive hash tables can be used to compute the k-nearest neighbors.

In SVMs, basis functions are centered on training samples; the training procedure selects a subset of the basis functions. The number of selected basis functions, and the number of training samples that have to be stored, is typically much smaller than the cardinality of the training dataset. SVMs build a linear decision boundary with the largest possible distance from the training samples. Only the closest points to the separators, the support vectors, are stored. To determine the parameters of SVMs, a nonlinear optimization problem with a convex objective function has to be solved, for which efficient algorithms exist. An important feature of SVMs is that by applying a kernel function they can embed data into a higher dimensional space, in which data points can be linearly separated. The kernel function measures the similarity between two points in the input space; it is expressed as the inner product of the input points mapped into a higher dimension feature space in which data become linearly separable. The simplest example is the linear kernel, in which the mapping function is the identity function. However, provided that we can express everything in terms of kernel evaluations, it is not necessary to explicitly compute the mapping in the feature space. Indeed, in the case of one of the most commonly used kernel functions,  the Gaussian kernel, the feature space has infinite dimensions.

\subsection{Unsupervised learning}
Social network analysis, genes clustering and market research are among the most successful applications of unsupervised learning methods.

In the case of unsupervised learning the training dataset consists only of a set of input vectors $\mathbf{x}$. While unsupervised learning can address different tasks, clustering or cluster analysis is the most common.

Clustering is the process of grouping data so that the intra-cluster similarity is high, while the inter-cluster similarity is low. The similarity is typically expressed as a distance function, which depends on the type of data. There exists a variety of clustering approaches. Here, we focus on two algorithms, k-means and Gaussian mixture model as examples of partitioning approaches and model-based approaches, respectively, given their wide area of applicability. The reader is referred to \cite{han2011data} for a comprehensive overview of cluster analysis.

k-means is perhaps the most well-known clustering algorithm (see chapter X in \cite{han2011data}). It is an iterative algorithm starting with an initial partition of the data into k clusters. Then the centre of each cluster is computed and data points are assigned to the cluster with the closest centre. The procedure - centre computation and data assignment - is repeated until the assignment does not change or a predefined maximum number of iterations is exceeded. Doing so, the algorithm may terminate at a local optimum partition. Moreover, k-means is well known to be sensitive to outliers. It is worth noting that there exists ways to compute k automatically \cite{hastie2009}, and an online version of the algorithm exists.

\begin{figure}
\centering
    \includegraphics[width=0.95\columnwidth]{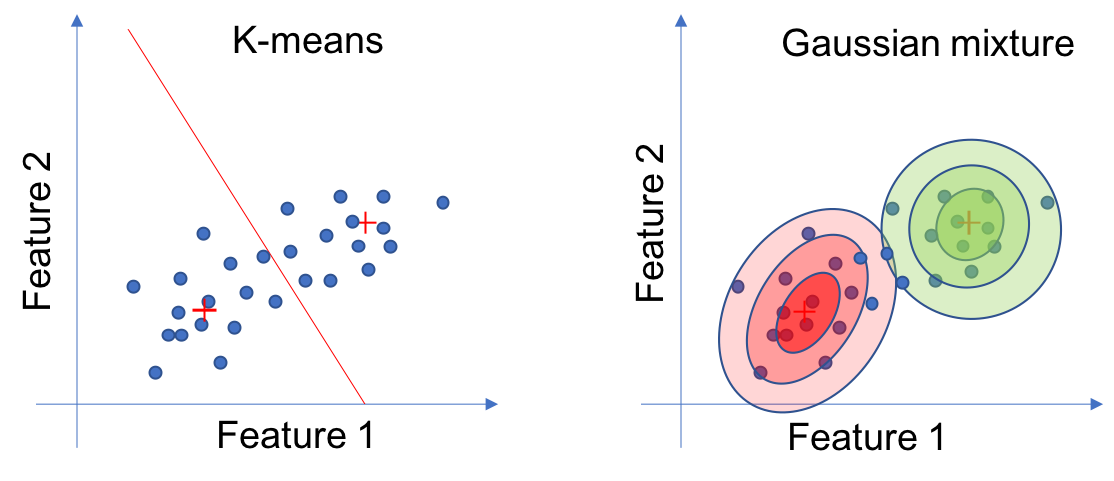}
    \caption{Difference between k-means and Gaussian mixture model clustering a given set of data samples.} \label{fig:means_mixture}
\end{figure}

While k-means assigns each point uniquely to one cluster, probabilistic approaches allow a soft assignment and provide a measure of the uncertainty associated with the assignment. Figure \ref{fig:means_mixture} shows the difference between k-means and a probabilistic Gaussian Mixture Model (GMM). GMM, a linear superposition of Gaussian distributions, is one of the most widely used probabilistic approaches to clustering. The parameters of the model are the mixing coefficient of each Gaussian component, the mean and the covariance of each Gaussian distribution. To maximize the log likelihood function with respect to the parameters given a dataset, the expectation maximization algorithm is used, since no closed form solution exists in this case. The initialization of the parameters can be done using k-means. In particular, the mean and covariance of each Gaussian component can be initialized to sample means and covariances of the cluster obtained by k-means, and the mixing coefficients can be set to the fraction of data points assigned by k-means to each cluster. After initializing the parameters and evaluating the initial value of the log likelihood, the algorithm alternates between two steps. In the expectation step, the current values of the parameters are used to determine the \lq\lq responsibility'' of each component for the observed data (i.e., the conditional probability of latent variables given the dataset). The maximization step uses these responsibilities to compute a maximum likelihood estimate of the model's parameters. Convergence is checked with respect to the log likelihood function or the parameters.

\subsection{Semi-supervised learning}
Semi-supervised learning methods are a hybrid of the previous two introduced above, and address problems in which most of the training samples are unlabeled, while only a few labeled data points are available. The obvious advantage is that in many domains a wealth of unlabeled data points is readily available. Semi-supervised learning is used for the same type of applications as supervised learning. It is particularly useful when labeled data points are not so common or too expensive to obtain and the use of available unlabeled data can improve performance.

\begin{figure}
\centering
    \includegraphics[width=0.95\columnwidth]{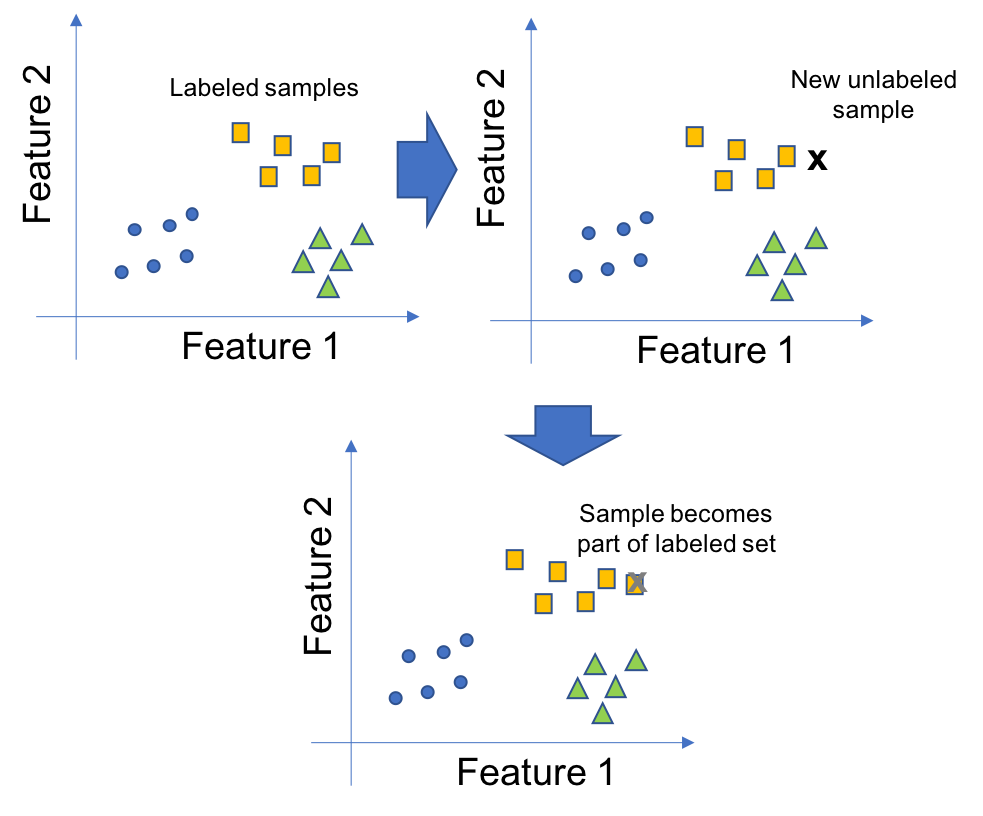}
    \caption{Sample step of the self-training mechanism, where an unlabeled point is matched against labeled data to become part of the labeled data set.} \label{fig:self_training}
\end{figure}

Self-training is the oldest form of semi-supervised learning \cite{chapelle2006}. It is an iterative process; during the first stage only labeled data points are used by a supervised learning algorithm. Then, at each step, some of the unlabeled points are labeled according to the prediction resulting for the trained decision function and these points are used along with the original labeled data to retrain using the same supervised learning algorithm. This procedure is shown in Fig. \ref{fig:self_training}.

Since the introduction of self-training, the idea of using labeled and unlabeled data has resulted in many semi-supervised learning algorithms. According to the classification proposed in \cite{chapelle2006},
semi-supervised learning techniques can be organized in four classes: i) methods based on generative models\footnote{Generative methods estimate the joint distribution of the input and output variables. From the joint distribution one can obtain the conditional distribution $p(\mathbf{y}|\mathbf{x})$, which is then used to predict the output values in correspondence to new input values. Generative methods can exploit both labeled and unlabeled data.}; ii) methods based on the assumption that the decision boundary should lie in a low-density region; iii) graph-based methods; iv) two-step methods (first an unsupervised learning step to change the data representation or construct a new kernel; then a supervised learning step based on the new representation or kernel).

\subsection{Reinforcement Learning}
Reinforcement Learning (RL) is used, in general, to address applications such as robotics, finance (investment decisions), inventory management, where the goal is to learn a policy, i.e., a mapping between states of the environment into actions to be performed, while directly interacting with the environment.

The RL paradigm allows agents to learn by exploring the available actions and refining their behavior using only an
evaluative feedback, referred to as the reward.
The agent's goal is to maximize its long-term performance. Hence,  the agent does not just take into account the immediate reward, but it evaluates the consequences of its actions on the future. Delayed reward and trial-and-error constitute the
two most significant features of RL.

RL is usually performed in the context of Markov decision processes (MDP). 
The agent's perception at time $k$ is represented as a state $\mathbf{s}_k \in \mathbf{S}$,
where $\mathbf{S}$ is the finite set of environment states. The agent interacts with the environment by performing actions. At time $k$ the agent selects an 
action $\mathbf{a}_k \in \mathbf{A}$, where $\mathbf{A}$ is the finite set of actions of the agent, which
could trigger a transition to a new state. The agent will receive
a reward as a result of the transition, according to the reward
function $\rho: \mathbf{S} \times \mathbf{A} \times \mathbf{S} \rightarrow R$. The agent’s goal is to find
the sequence of state-action pairs that maximizes
the expected discounted reward, i.e., the optimal policy. In the context of MDP, it has
been proved that an optimal deterministic and stationary policy
exists. There exist a number of algorithms that learn the optimal policy 
both in case the state transition and reward functions are known (model-based learning) and in  case they are not (model-free learning). The most used RL algorithm is  Q-learning, a model-free algorithm that estimates the  optimal
action-value function (see chapter VI in \cite{sutton1998reinforcement}). An action-value function, named Qfunction, is the expected return of a state-action pair for a given
policy. The optimal action-value function, $Q*$, corresponds to
the maximum expected return for a state-action pair. After learning function $Q*$, the agent selects the action with the corresponding highest Q-value in correspondence to the current state.

A table-based solution such as the one described above is only suitable in case of problems with limited state-action space. In order to generalize the policy learned in correspondence to states not previously experienced by the agent, RL methods can be combined with existing function approximation methods, e.g., neural networks.

\subsection{Overfitting, underfitting and model selection}

In this section, we discuss a well-known problem of ML algorithms along with its solutions. Although we focus on supervised learning techniques, the discussion is also relevant for unsupervised learning methods.

\begin{figure}
\centering
    \includegraphics[width=0.95\columnwidth]{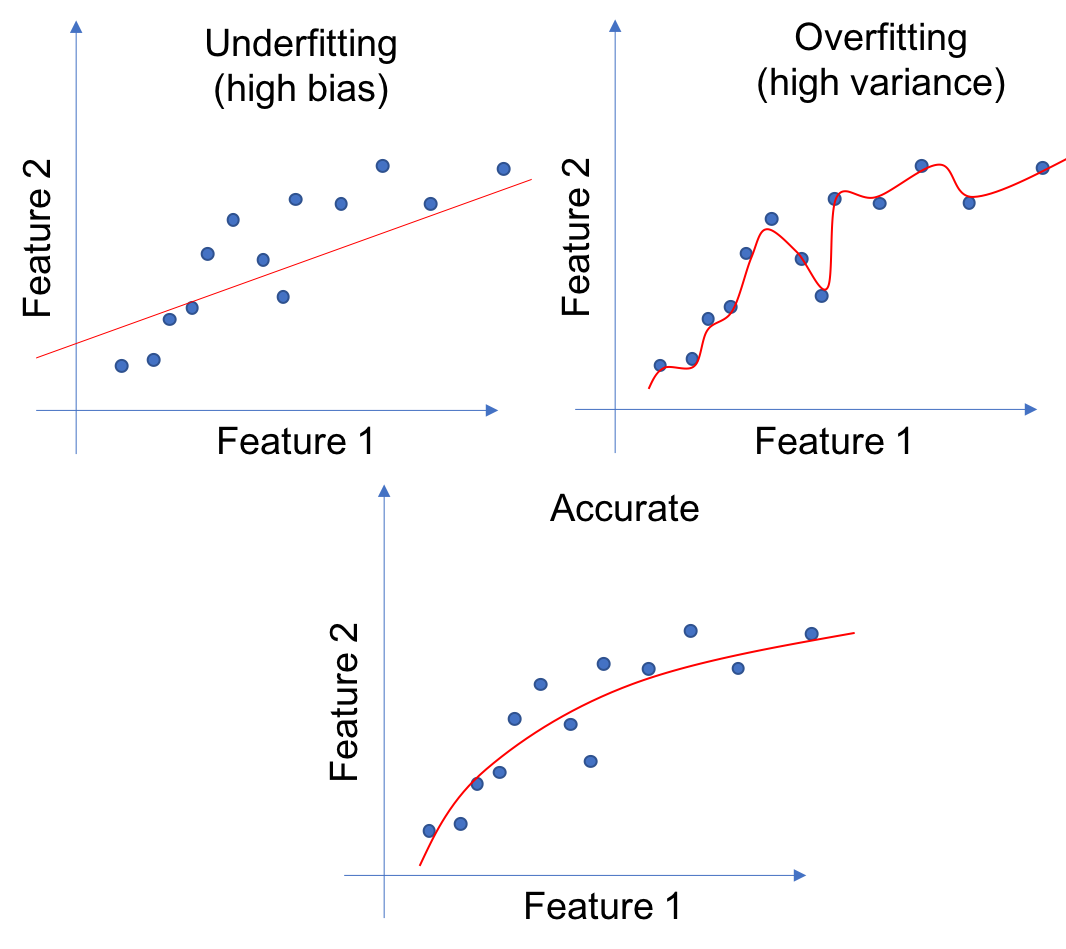}
    \caption{Difference between underfitting and overfitting.} \label{fig:fitting}
\end{figure}

Overfitting and underfitting are two sides of the same coin: model selection. Overfitting happens when the model we use is too complex for the available dataset (e.g., a high polynomial order in the case of linear regression with polynomial basis functions or a too large number of hidden neurons for a neural network). In this case, the model will fit the training data too closely\footnote{As an extreme example, consider a simple regression problem for predicting a real-value target variable as a function of a real-value observation variable. Let us assume a linear regression model with polynomial basis function of the input variable. If we have $N$ samples and we select $N$ as the order of the polynomial, we can fit the model perfectly to the data points.}, including noisy samples and outliers, but will result in very poor generalization, i.e., it will provide inaccurate predictions for new data points. At the other end of the spectrum,  underfitting is caused by the selection of models that are not complex enough to capture important features in the data (e.g., when we use a linear model to fit quadratic data). Fig. \ref{fig:fitting} shows the difference between underfitting and overfitting, compared to an accurate model.

Since the error measured on the training samples is a poor indicator for generalization, to evaluate the model performance the available dataset is split into two, the training set and the test set. The model is trained on the training set and then evaluated using the test set. Typically around $70\%$ of the samples are assigned to the training set and the remaining $30\%$ are assigned to the test set. Another option that is very useful in case of a limited dataset is to use cross-validation so that as much of the available data as possible is exploited for training. In this case, the dataset is divided into $k$ subsets. The model is trained $k$ times using each of the $k$ subset for validation and the remaining $(k-1)$ subsets for training. The performance is averaged over the $k$ runs. In case of overfitting, the error measured on the test set is high and the error on the training set is small. On the other hand, in the case of underfitting,  both the error measured on the training set and the test set are usually high. 

There are different ways to select a model that does not exhibit overfitting and underfitting. One possibility is to train a range of models, compare their performance on an independent dataset (the validation set), and then select the one with the best performance. However, the most common technique is regularization. It consists of adding an extra term - the regularization term - to the error function used in the training stage. The simplest form of the regularization term is the sum of the squares of all parameters, which is known as weight decay and drives parameters towards zero.  Another common choice is the sum of the absolute values of the parameters (lasso). An additional parameter, the regularization coefficient $\lambda$, weighs the relative importance of the regularization term and the data-dependent error. A large value of $\lambda$ heavily penalizes large absolute values of the parameters. It should be noted that the data-dependent error computed over the training set increases with $\lambda$. The error computed over the validation set is high for both small and high $\lambda$ values. In the first case, the regularization term has little impact potentially resulting in overfitting. In the latter case, the data-dependent error has little impact resulting in a poor model performance. A simple automatic procedure for selecting the best   $\lambda$ consists of training the model with a range of values for the regularization parameter and select the value that corresponds to the minimum validation error. In the case of NNs with a large number of hidden units, dropout - a technique that consists of randomly removing units and their connections during training - has been shown to outperform other regularization methods    \cite{srivastava2014dropout}.
 

%% file: sections/Optical_Networks_and_Systems_Problems_and_their_Mapping_into_ML_problems.tex
\section{Motivation for using machine learning in optical networks and Systems}
\label{sec:3}

\begin{figure*}
\centering
    \includegraphics[width=18cm]{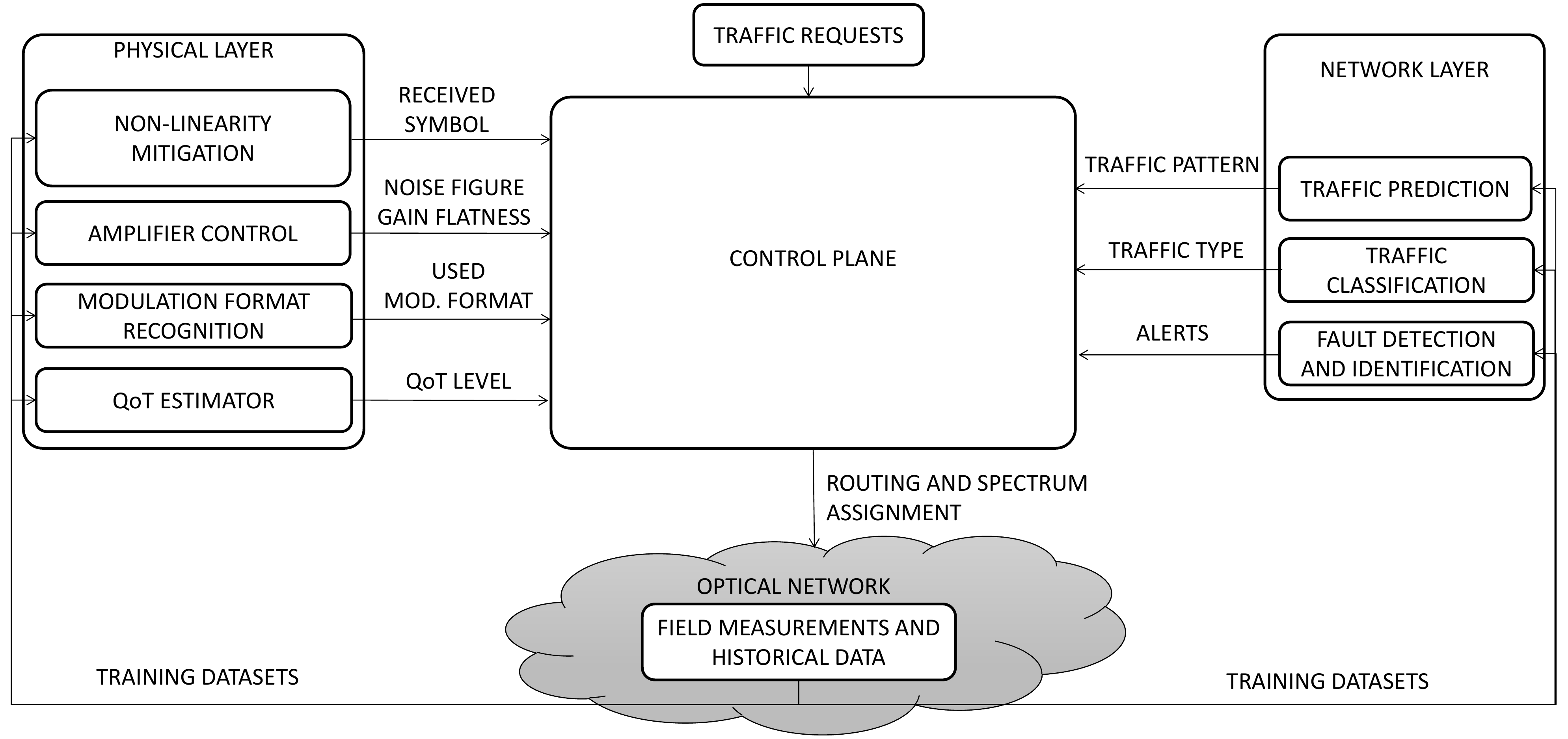}
    \caption{The general framework of a ML-assisted optical network.} \label{fig:framework}
\end{figure*}

In the last few years, the application of mathematical approaches derived from the ML discipline have attracted the attention of many researchers and practitioners in the optical communications and networking fields. In a general sense, the underlying motivations for this trend can be identified as follows:
\begin{itemize}
\item \textit{increased system complexity}: the adoption of advanced transmission techniques, such as those enabled by coherent technology \cite{huawei}, and the introduction of extremely flexible networking principles, such as, e.g., the EON paradigm, have made the design and operation of optical networks extremely complex, due to the high number of tunable parameters to be considered (e.g., modulation formats, symbol rates, adaptive coding rates, adaptive channel bandwidth, etc.); in such a scenario, accurately modeling the system through closed-form formulas is often very hard, if not impossible, and in fact \lq\lq margins'' are typically adopted in the analytical models, leading to resource underutilization and to consequent  increased system cost; on the contrary, ML methods can capture complex non-linear system behaviour with relatively simple training of supervised and/or unsupervised algorithms which exploit knowledge of historical network data, and therefore to solve complex cross-layer problems, typical of the optical networking field;

\item \textit{increased data availability}: modern optical networks are equipped with a large number of monitors, able to provide several types of information on the entire system, e.g., traffic traces, signal quality indicators (such as BER), equipment failure alarms, users' behaviour etc.; here, the enhancement brought by ML consists of \textit{simultaneously} leveraging the plethora of collected data and discover hidden relations between various types of information.

\end{itemize}

The application of ML to physical layer use cases is mainly motivated by the presence of non-linear effects in optical fibers, which make analytical models inaccurate or even too complex.
This has implications, e.g., on the performance predictions of optical communication systems, in terms of BER, quality factor (Q-factor) and also for signal demodulation \cite{Wass:17},\cite{Zibar:12}, \cite{Zibar2016}. 

Moving from the physical layer to the networking layer, the same motivation applies for the application of ML techniques. 
In particular, design and management of optical networks is continuously evolving, driven by the enormous increase of transported traffic and drastic changes in traffic requirements, e.g., in terms of capacity, latency, user experience and Quality of Service (QoS). Therefore, current optical networks are expected to be run at much higher utilization than in the past, while providing strict guarantees on the provided quality of service.  While aggressive optimization and traffic-engineering methodologies are required to achieve these objectives, such complex methodologies may suffer scalability issues, and involve unacceptable  computational complexity. In this context, ML  is regarded as a promising methodological area to address this issue, as it enables automated network self-configuration and fast decision-making by leveraging the plethora of data that can be retrieved via network monitors, and allowing network engineers to build data-driven models for more accurate and optimized network provisioning and management.

Several use cases can benefit from the application of ML and data analytics techniques. In this paper we divide these use cases in $i)$ \textit{physical layer} and $ii)$ \textit{network layer} use cases.  The remainder of this section provides a high-level introduction to the main applications of ML in optical networks, as graphically shown in Fig. \ref{fig:framework}, and motivates why ML can be beneficial in each case. A detailed survey of existing studies is then provided in Sections \ref{sec:4} and \ref{sec:5}, for physical layer and network layer use cases, respectively.

\subsection{Physical layer domain}

As mentioned in the previous section, several challenges need to be addressed at the physical layer of an optical network, typically to evaluate the performance of the transmission system and to check if any signal degradation influences existing lightpaths. Such monitoring can be used, e.g., to trigger proactive procedures, such as tuning of launch power, controlling gain in optical amplifiers, varying modulation format, etc., before irrecoverable signal degradation occurs.
In the following, a description of the applications of ML at the physical layer is presented.

\begin{itemize}

\item \textit{QoT estimation}. \newline
Prior to the deployment of a new lightpath, a system engineer needs to estimate the Quality of Transmission (QoT) for the new lightpath, as well as for the already existing ones.
The concept of Quality of Transmission generally refers to a number of physical layer parameters, such as received Optical Signal-to-Noise Ratio (OSNR), BER, Q-factor, etc., which have an impact on the \lq\lq readability'' of the optical signal at the receiver. Such parameters give a quantitative measure to check if a pre-determined level of QoT would be guaranteed, and are affected by several tunable design parameters, such as, e.g., modulation format, baud rate, coding rate, physical path in the network, etc. Therefore, optimizing this choice is not trivial and often this large variety of possible parameters challenges the ability of a system engineer to address manually all the possible combinations of lightpath deployment. 

As of today, existing (pre-deployment) estimation techniques for lightpath QoT belong to two categories: 1) \lq\lq exact'' analytical models estimating physical-layer impairments, which provide accurate results, but incur heavy computational requirements and 2) marginated formulas, which are computationally faster, but typically introduce high marginations that lead to underutilization of network resources. Moreover, it is worth noting that, due to the complex interaction of multiple system parameters (e.g., input signal power, number of channels, link type, modulation format, symbol rate, channel spacing, etc.) and, most importantly, due to the nonlinear signal propagation through the optical channel, deriving accurate analytical models is a challenging task, and assumptions about the system under consideration must be made in order to adopt approximate models.
Conversely, ML constitutes a promising means to automatically predict whether unestablished lightpaths will meet the required system QoT threshold. 

\textbf{Relevant ML techniques}:  ML-based classifiers can be trained using supervised learning\footnote{Note that, specific solutions adopted in literature for QoT estimation, as well as for other physical- and network-layer use cases, will be detailed in the literature surveys provided in Sections \ref{sec:4} and \ref{sec:5}.} to create direct input-output relationship between QoT observed at the receiver and corresponding lightpath configuration in terms of, e.g., utilized modulation format, baud rate and/or physical route in the network.

\item \textit{Optical amplifiers control}. \newline
In current optical networks, lightpath provisioning is becoming more dynamic, in response to the emergence of new services that require huge amount of bandwidth over limited periods of time.
Unfortunately, dynamic set-up and tear-down of lightpaths over different wavelengths forces network operators to reconfigure network devices \lq\lq on the fly'' to maintain physical-layer stability. In response to rapid changes of lightpath deployment, Erbium Doped Fiber Amplifiers (EDFAs) suffer from wavelength-dependent power excursions.
Namely, when a new lightpath is established (i.e., added) or when an existing lightpath is torn down (i.e., dropped), the discrepancy of signal power levels between different channels (i.e., between lightpaths operating at different wavelengths) depends on the specific wavelength being added/dropped into/from the system.
Thus, an automatic control of pre-amplification signal power levels is required, especially in case a cascade of multiple EDFAs is traversed, to avoid that excessive post-amplification power discrepancy between different lightpaths may cause signal distortion.

\textbf{Relevant ML techniques}: Thanks to the availability of historical data retrieved by monitoring network status, ML regression algorithms can be trained to accurately predict post-amplifier power excursion in response to the add/drop of specific wavelengths to/from the system.

\item  \textit{Modulation format recognition (MFR)}. \newline
Modern optical transmitters and receivers provide high flexibility in the utilized bandwidth, carrier frequency and modulation format, mainly to adapt the transmission to the required bit-rate and optical reach in a flexible/elastic networking environment. Given that at the transmission side an arbitrary coherent optical modulation format can be adopted, knowing this decision in advance also at the receiver side is not always possible, and this may affect proper signal demodulation and, consequently, signal processing and detection. 

\textbf{Relevant ML techniques}: Use of supervised ML algorithms can help the modulation format recognition at the receiver, thanks to the opportunity to learn the mapping between the adopted modulation format and the features of the incoming optical signal.

\item  \textit{Nonlinearity mitigation}. \newline
Due to optical fiber nonlinearities, such as Kerr effect, self-phase modulation (SPM) and cross-phase modulation (XPM), the behaviour of several performance parameters, including BER, Q-factor, Chromatic Dispersion (CD), Polarization Mode Dispersion (PMD), is highly unpredictable, and this may cause signal distortion at the receiver (e.g., I/Q imbalance and phase noise).
Therefore, complex analytical models are often adopted to react to signal degradation and/or compensate undesired nonlinear effects.

\textbf{Relevant ML techniques}: While approximated analytical models are usually adopted to solve such complex non-linear problems, supervised ML models can be designed to directly capture the effects of such nonlinearities, typically exploiting knowledge of historical data and creating input-output relations between the monitored parameters and the desired outputs.

\item  \textit{Optical performance monitoring (OPM)}. \newline
With increasing capacity requirements for optical communication systems, performance monitoring is vital to ensure robust and reliable networks. Optical performance monitoring aims at estimating the transmission parameters of the optical fiber system, such as BER, Q-factor, CD, PMD, during lightpath lifetime. Knowledge of such parameters can be then utilized to accomplish various tasks, e.g., activating polarization compensator modules, adjusting launch power, varying the adopted modulation format, re-route lightpaths, etc. Typically, optical performance parameters need to be collected at various monitoring points along the lightpath, thus large number of monitors are required, causing increased system cost. Therefore, efficient deployment of optical performance monitors in the proper network locations is needed to extract network information at reasonable cost.

\textbf{Relevant ML techniques}:  To reduce the amount of monitors to deploy in the system, especially at intermediate points of the lightpaths, supervised learning algorithms can be used to learn the mapping between the optical fiber channel parameters and the properties of the detected signal at the receiver, which can be retrieved, e.g., by observing statistics of power eye diagrams, signal amplitude, OSNR, etc.

\end{itemize}

\subsection{Network layer domain}

At the network layer, several other use cases for ML arise. Provisioning of new lightpaths or restoration of existing ones upon network failure require complex and fast decisions that depend on several quickly-evolving data, since, e.g., operators must take into consideration the impact onto existing connections provided by newly-inserted traffic. In general, an estimation of users' and service requirements is desirable for an effective network operation, as it allows to avoid over-provisioning of network resources and to deploy resources with adequate margins at a reasonable cost. We identify the following main use cases.

\begin{itemize}

\item \textit{Traffic prediction}. \newline
Accurate traffic prediction in the time-space domain allows operators to effectively plan and operate their networks. In the design phase, traffic prediction allows to reduce over-provisioning as much as possible.  During network operation, resource utilization can be optimized by performing traffic engineering based on real-time data, eventually re-routing existing traffic and reserving resources for future incoming traffic requests. 

\textbf{Relevant ML techniques}: Through knowledge of historical data on users' behaviour and traffic profiles in the time-space domain, a supervised learning algorithm can be trained to predict future traffic requirements and consequent resource needs. This allows network engineers to activate, e.g., proactive traffic re-routing and periodical network re-optimization so as to accommodate all users traffic and simultaneously reduce network resources utilization.

Moreover, unsupervised learning algorithms can be also used to extract common traffic patterns in different portions of the network. Doing so, similar design and management procedures (e.g., deployment and/or reservation of network capacity) can be activated also in different parts of the network, which instead show similarities in terms of traffic requirements, i.e., belonging to a same traffic profile \textit{cluster}.

Note that, application of traffic prediction, and the relative ML techniques, vary substantially according to the considered network segment (e.g., approaches for intra-datacenter networks may be different than those for access networks), as traffic characteristics strongly depend on the considered network segment.

\item \textit{Virtual topology design (VTD) and reconfiguration}. \newline
The abstraction of communication network services by means of a virtual topology is widely adopted by network operators and service providers. This abstraction consists of representing the connectivity between two end-points (e.g., two data centers) via an adjacency in the virtual topology, (i.e., a virtual link), although the two end-points are not necessarily physically connected. After the set of all virtual links has been defined, i.e., after all the lightpath requests have been identified, VTD requires solving a Routing and Wavelength Assignment (RWA) problem for each lightpath on top of the underlying physical network. Note that, in general, many virtual topologies can co-exist in the same physical network, and they may represent, e.g., service required by different customers, or even different services, each with a specific set of requirements (e.g., in terms of QoS, bandwidth, and/or latency), provisioned to the same customer. 

VTD is not only necessary when a new service is provisioned and new resources are allocated in the network. In some cases, e.g., when network failures occur or when the utilization of network resources undergoes re-optimization procedures, \textit{existing} (i.e., already-designed) virtual topologies shall be rearranged, and in these cases we refer to the \textit{VT reconfiguration}.

To perform design and reconfiguration of virtual topologies, network operators not only need to provision (or reallocate) network capacity for the required services, but may also need to provide additional resources according to the specific service characteristics, e.g., for guaranteeing service protection and/or meeting QoS or latency requirements. This type of service provisioning is often referred to as \textit{network slicing}, due to the fact that each provisioned service (i.e., each VT) represents a \textit{slice} of the overall network.

\textbf{Relevant ML techniques}: To address VTD and VT reconfiguration, ML classifiers can be trained to optimally decide how to allocate network resources, by simultaneously taking into account a large number of different and heterogeneous service requirements for a variety of virtual topologies (i.e., \textit{network slices}), thus enabling fast decision making and optimized resources provisioning, especially under dynamically-changing network conditions.


\item \textit{Failure management}. \newline
When managing a network, the ability to perform failure detection and localization or even to determine the cause of network failure is crucial as it may enable operators to promptly perform traffic re-routing, in order to maintain service status and meet Service Level Agreements (SLAs), and rapidly recover from the failure. 
Handling network failures can be accomplished at different levels. E.g., performing failure detection, i.e., identifying the set of lightpaths that were affected by a failure, is a relatively simple task, which allows network operators to only reconfigure the affected lightpaths by, e.g., re-routing the corresponding traffic. Moreover, the ability of performing also failure localization enables the activation of recovery procedures. This way, pre-failure network status can be restored, which is, in general, an optimized situation from the point of view of resources utilization. Furthermore, determining also the cause of network failure, e.g., temporary traffic congestion, devices disruption, or even anomalous behaviour of failure monitors, is useful to adopt the proper restoring and traffic reconfiguration procedures, as sometimes remote reconfiguration of lightpaths can be enough to handle the failure, while in some other cases in-field intervention is necessary. Moreover, prompt identification of the failure cause enables fast equipment repair and consequent reduction in Mean Time To Repair (MTTR).

\textbf{Relevant ML techniques}:  ML can help handling the large amount of information derived from the continuous activity of a huge number of network monitors and alarms. E.g., ML classifiers algorithms can be trained to distinguish between regular and anomalous (i.e., degraded) transmission. Note that, in such cases, semi-supervised approaches can be also used, whenever labeled data are scarce, but a large amount of unlabeled data is available. 
Further, ML classifiers can be trained to distinguish failure causes, exploiting the knowledge of previously observed failures.

\item \textit{Traffic flow classification}. \newline
When different types of services coexist in the same network infrastructure, classifying the corresponding traffic flows before their provisioning may enable efficient resource allocation, mitigating the risk of under- and over-provisioning. Moreover, accurate flow classification is also exploited for already provisioned services to apply flow-specific policies, e.g., to handle packets priority, to perform flow and congestion control, and to guarantee proper QoS to each flow according to the SLAs. 

\textbf{Relevant ML techniques}: Based on the various traffic characteristics and exploiting the large amount of information carried by data packets, supervised learning algorithms can be trained to extract hidden traffic characteristics and perform fast packets classification and flows differentiation.

\item \textit{Path computation}. \newline
When performing network resources allocation for an incoming service request, a proper path should be selected in order to efficiently exploit the available network resources to accommodate the requested traffic with the desired QoS and without affecting the existing services, previously provisioned in the network. Traditionally, path computation is performed by using cost-based routing algorithms, such as Dijkstra, Bellman-Ford, Yen algorithms, which rely on the definition of a pre-defined cost metric (e.g., based on the distance between source and destination, the end-to-end delay, the energy consumption, or even a combination of several metrics) to discriminate between alternative paths. 

\textbf{Relevant ML techniques}: In this context, use of supervised ML can be helpful as it allows to simultaneously consider several parameters featuring the incoming service request together with current network state information and map this information into an optimized routing solution, with no need for complex network-cost evaluations and thus enabling fast path selection and service provisioning.

\end{itemize}

\subsection{A bird-eye view of the surveyed studies}

The physical- and network-layer use cases described above have been tackled in existing studies by exploiting several ML tools (i.e., supervised and/or unsupervised learning, etc.) and leveraging different types of network monitored data (e.g., BER, OSNR, link load, network alarms, etc.).

In Tables \ref{tab:problems_PHY} and \ref{tab:problems_NET} we summarize the various physical- and network-layer use cases and highlight the features of the ML approaches which have been used in literature to solve these problems. In the tables we also indicate specific reference papers addressing these issues, which will be described in the following sections in more detail. Note that another recently published survey \cite{MATA201843} proposes a very similar categorization of existing applications of artificial intelligence in optical networks.

\begin{table*}[h!]
	\caption{Different use cases at physical layer and their characteristics.}
	\centering
	\begin{tabular}{p{1.7cm} l p{2.5cm} p{3.5cm} p{3cm} p{1.5cm} r}
		\toprule
		Use Case	&ML category &ML methodology		&Input data &Output data	&Training data		&Ref.\\
		\midrule
QoT estimation	&supervised	 	&kriging, $L_2$-norm minimization 	&OSNR (historical data)			&OSNR	&synthetic 	&\cite{angelou2012optimized}\\
				& 	 			& 									&OSNR/Q-factor	&BER					&synthetic 	&\cite{sartzetakis2016quality,pointurier2011cross}\\
                & 	 			& 									&OSNR/PMD/CD/SPM	&blocking prob.		&synthetic 		&\cite{sambo2010lightpath}\\
                
                &	 			&CBR						 		&error vector magnitude, OSNR	&Q-factor	&real 		&\cite{caballero2012experimental}\\
                &	 			&									&lightpath route, length, number of co-propagating lightpaths	&Q-factor	&synthetic 		&\cite{jimenez2013cognitive,de2013cognitive}\\
                
                &	 			&RF 								&lightpath route, length, MF, traffic volume	&BER	&synthetic 		&\cite{barletta17QoT}\\
                
                &	 			&regression 	&SNR (historical data)	&SNR	&synthetic 		&\cite{seve2017learning}\\
                
                &	 			&NN								&lightpath route and length, number of traversed
EDFAs, degree of destination, used channel wavelength	&Q-factor 	&synthetic 		&\cite{panayiotou2016data,panayiotou2017performance}\\ 
&	 			&k-nearest neighbor, RF, SVM								&total link length, span length, channel launch power, MF and data rate	&BER	&	synthetic	&\cite{Aladin2018}\\
&	 			&NN								&channel loadings and launch power settings	&Q-factor	&real	&\cite{WMoQoT2018}\\
&	 			&NN								& source-destination nodes, link occupation, MF, path length, data rate	& BER	&	real	&\cite{Proietti2018}\\

\midrule
OPM		&supervised		&NN 			&eye diagram and amplitude histogram param.							&OSNR/PMD/CD	&real 		&\cite{tanimura2016osnr}\\
		&				&NN, SVM		&asynchronous amplitude histogram 									&MF				&real 		&\cite{Thrane2017}\\
		&	 			&NN 			&asyncrhonous constellation diagram and amplitude histogram param.	&OSNR/PMD/CD 	&synthetic 	&\cite{wu2009applications,jargon2010optical,shen2010optical,khan2012optical}\\
       &			&Kernel-based ridge regression 					&eye diagram and phase portraits param.	&PMD/CD	&real 		&\cite{anderson2009multi}\\
       &			&NN 					&Horizontal and Vertical polarized I/Q samples from ADC			&OSNR, MF, symbol rate	&real 		&\cite{Tanimura2018}\\
       &			&Gaussian Processes		&monitoring data (OSNR vs $\lambda$)							&Q-factor	&real 		&\cite{Meng2018}\\
\midrule        
Optical amplifiers control	&supervised &CBR	&power mask param. (NF, GF)	&OSNR	&real 	&\cite{moura2016cognitive,oliveira2013demonstration}\\
                  			&			&NNs		&EDFA input/output power	&EDFA operating point	&real 	&\cite{barboza2013self,bastos2013mapping}\\
         					&			&Ridge regression, Kernelized Bayesian regr.		&WDM channel usage	&post-EDFA power discrepancy	&real 	&\cite{huang2017dynamic}\\
                            &unsupervised &evolutional alg.		&EDFA input/output power		&EDFA operating point	&real 	&\cite{de2013edfa}\\

\midrule
MF recognition		&unsupervised	&6 clustering alg.		&Stokes space param.					&MF	&synthetic 	&\cite{boada2015clustering}\\
					&	 			&k-means				&received symbols						&MF	&real 		&\cite{gonzalez2010cognitive}\\
					&supervised		&NN						&asynchronous amplitude histogram		&MF	&synthetic 	&\cite{zhang2016modulation}\\
					&				&NN, SVM				&asynchronous amplitude histogram 		&MF	&real 		&\cite{khan2012modulation},\cite{khan2016modulation},\cite{Thrane2017}\\
                    &				&variational Bayesian techn. for GMM	&Stokes space param.	&MF	&real 	&\cite{borkowski2013stokes}\\

\midrule
Non-linearity mitigation		&supervised	&Bayesian filtering, NNs, EM	&received symbols	&OSNR, Symbol error rate	&real 		&\cite{Zibar:12,Zibar2016,Zibar2016a}\\
					&		&ELM									&received symbols	&self-phase modulation		&synthetic 		&\cite{shen2011fiber}\\
					&		&k-nearest neighbors					&received symbols	&BER		&real 		&\cite{wang2016nonlinearity}\\
					&		&Newton-based SVM						&received symbols	&Q-factor	&real 		&\cite{giacoumidis2017reduction}\\
&					&binary SVM										&received symbols	&symbol decision boundaries	&synthetic 		&\cite{wang2015nonlinear}\\
& 					&NN 											&received subcarrier symbols		&Q-factor	&synthetic 		&\cite{jarajreh2015artificial}\\
& 					&GMM 								&post-equalized symbols	& decoded symbols with impairment estimated and/or mitigated	&real	&\cite{Lu2018}\\
& 					&Clustering 				&received constellation with nonlinearities	&nonlinearity mitigated constellation points 	&real	&\cite{XLu2018}\\
            & 	&NN		&sampled received signal sequences 	&equalized signal with reduced ISI 	&real 		&\cite{Li2018, Chuang2018, Pli2018, Jones2018, Hager2018, SLiu2018}\\
	&unsupervised	&k-means	&received constellation	&density-based spatial constellation clusters and their optimal centroids	&real 		&\cite{Zhang2018}\\

		\bottomrule
	\end{tabular}
	\label{tab:problems_PHY}
\end{table*}


 \begin{table*}[h!]
	\caption{Different use cases at network layer and their characteristics.}
	\centering
	\begin{tabular}{p{2cm} l p{2.8cm} p{3.5cm} p{3cm} p{1.5cm} r}
		\toprule
		Use Case	&ML category &ML methodology		&Input data &Output data	&Training data		&Ref.\\
		
\midrule
Traffic prediction and virtual topology (re)design		&supervised	&ARIMA			&historical real-time traffic matrices	&predicted traffic matrix	&synthetic 	&\cite{Fernandez2013}, \cite{Fernandez2015}\\
								&  				&NN									&historical end-to-end maximum bit-rate traffic	&predicted end-to-end traffic	&synthetic  &\cite{Morales2016}, \cite{Morales2017}\\\\
                                
		&				&Reinforcement learning								&previous solutions of a multi-objective GA for VTD	&updated VT		&synthetic			&\cite{Fernandez2012Energy}, \cite{Fernandez2012Surv}\\\\
        		&				&Recurrent NN								&	historical aggregated traffic at different BBU pools	& predicted BBU pool traffic	&	real	& \cite{WMo2018}\\\\
                		&				& NN								&	historical traffic in intra-DC network	& predicted intra-DC traffic	&	real	& \cite{Yu2018}\\
                                &unsupervised 	&NMF, clustering					&CDR, PoI matrix	&similarity patterns in base station traffic	&real  &\cite{Troia2017}\\

\midrule 
Failure management				&supervised		&Bayesian Inference					&BER, received power	&list of failures for all lightpaths	&real 	&\cite{ruiz2016service}\\\\

                      			&		&Bayesian Inference, EM				&FTTH network dataset with missing data &complete dataset		&real 		&\cite{Tembo2016}, \cite{Gosselin2017}\\\\
                                
								&		&Kriging							&previously established lightpaths with already available failure localization and monitoring data	&estimate of failure localization at link level for all lightpaths	&real 	&\cite{christodoulopoulos2016exploiting}\\\\
                                &		&(1) LUCIDA: Regression and classification\newline (2) BANDO: Anomaly Detection	&	(1) LUCIDA: historic BER and received power, notifications from BANDO \newline(2) BANDO: maximum BER, threshold BER at set-up, monitored BER 		&(1) LUCIDA: failure classification \newline (2) BANDO: anomalies in BER 	&real 	&\cite{Vela_JLT_failures}\\\\
                                &		&Regression, decision tree, SVM		&BER, frequency-power pairs	&localized set of failures	&real 	&\cite{Vela:18}\\
								&		&SVM, RF, NN						&BER 						&set of failures	&real 	&\cite{Shahkarami2018}\\
								&		&regression and NN					&optical power levels, amplifier gain, shelf temperature, current draw, internal optical power	&detected faults	&real 	&\cite{Rafique2018}\\

\midrule
Flow classification				&supervised		&HMM, EM					&packet loss data		&loss classification: congestion-loss or contention-loss	&synthetic 	&\cite{Jayaraj2008OBS_TCP}\\\\
								&				&NN							&source/destination IP addresses, source/destination ports, transport layer protocol, packet sizes, and a set of intra-flow timings within the first 40 packets of a flow 		&classified flow for DC		&synthetic 	&\cite{Viljoen2016}\\\\

\midrule
Path computation				&supervised 	&Q-Learning					&traffic requests, set of candidate paths between each source-destination pair				&optimum paths for each source-destination pair to minimize burst-loss probability		&synthetic 	&\cite{Kiran2007OBS}\\\\
								&unsupervised	&FCM						&traffic requests, path lengths, set of modulation formats, OSNR, BER	&mapping of an optimum modulation format to a lightpath	&synthetic 	&\cite{Tronco2016}\\\\
		\bottomrule
	\end{tabular}
	\label{tab:problems_NET}
\end{table*}

%% file: sections/Detailed_survey_of_ML_in_Physical_Layer_Problems.tex
\section{Detailed survey of machine learning in physical layer domain}
\label{sec:4}

\subsection{Quality of Transmission estimation}
QoT estimation consists of computing transmission quality metrics such as OSNR, 
BER, Q-factor, CD or PMD based on measurements directly collected from the field by means of optical
performance monitors installed at the receiver side \cite{orchestra} and/or on lightpath characteristics. QoT estimation is typically applied in two scenarios:
\begin{itemize}
\item predicting the transmission quality of unestablished lightpaths based on historical observations and measurements collected from already deployed ones;
\item monitoring the transmission quality of already-deployed lightpaths with the aim of identifying faults and malfunctions.
\end{itemize}

QoT prediction of unestablished lightpaths relies on intelligent tools, capable of predicting whether a candidate lightpath will meet the required quality of service guarantees (mapped onto OSNR, BER or Q-factor threshold values): the problem is typically formulated as a binary classification problem, where the classifier outputs a yes/no answer based on the lightpath characteristics (e.g., its length, number of links, modulation format used for transmission, overall spectrum occupation of the traversed links etc.).

In \cite{jimenez2013cognitive} a cognitive Case Based Reasoning (CBR) approach is proposed, which relies on the maintenance of a knowledge database where information on the measured Q-factor of deployed lightpaths is stored, together with their route, selected wavelength, total length, total number and standard deviation of the number of co-propagating lightpaths per link. Whenever a new traffic requests arrives, the most \lq\lq similar'' one (where similarity is computed by means of the Euclidean distance in the multidimensional space of normalized features) 
is retrieved from the database and a decision is made by comparing the associated Q-factor measurement with a predefined system threshold. As a correct dimensioning and maintenance of the database greatly affect the performance of the CBR technique, algorithms
are proposed to keep it up to date and to remove old or useless entries. The trade-off between database size, computational time and effectiveness of the classification performance is extensively studied: in \cite{de2013cognitive}, the technique is shown to outperform state-of-the-art ML algorithms such as Naive Bayes, J48 tree and Random Forests (RFs). Experimental results achieved with data obtained from a real testbed are discussed in \cite{caballero2012experimental}.

A database-oriented approach is proposed also in \cite{seve2017learning} to reduce uncertainties on network parameters and design margins, where field data are collected by a software defined network controller and stored in a central repository. Then, a QTool is used to produce an estimate of the field-measured Signal-to-Noise Ratio (SNR) based on educated guesses on the (unknown) network parameters and such guesses are iteratively updated by means of a gradient descent algorithm, until the difference between the estimated and the field-measured SNR falls below a predefined threshold. The new estimated parameters are stored in the database and yield to new design margins, which can be used for future demands. The trade-off between database size and ranges of the SNR estimation error are evaluated via numerical simulations.

Similarly, in the context of multicast transmission in optical network, a NN is trained in \cite{panayiotou2016data,panayiotou2017performance,WMoQoT2018,Proietti2018} using as features the lightpath total length, the number of traversed EDFAs, the maximum link length, the degree of destination node and the channel wavelength used for transmission of candidate lightpaths, to predict whether the Q-factor will exceed a given system threshold. The NN is trained online with data mini-batches, according to the network evolution, to allow for sequential updates of the prediction model. A dropout technique is adopted during training to avoid overfitting. The classification output is exploited by a heuristic algorithm for dynamic routing and spectrum assignment, which decides whether the request must be served or blocked. The algorithm performance is assessed in terms of blocking probability.

\begin{figure}
\centering
    \includegraphics[width=9cm]{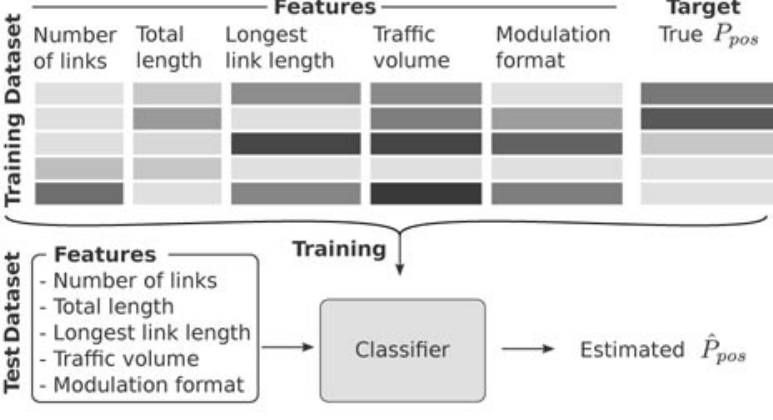}
    \caption{The classification framework adopted in \cite{barletta17QoT}.} \label{fig:classifier}
\end{figure}

A random forest binary classifier is adopted in \cite{barletta17QoT} to predict the probability that the BER of unestablished lightpaths will exceed a system threshold. As depicted in Figure \ref{fig:classifier}, the classifier takes as input a set of features including the total length and maximum link length of the candidate lightpath, the number of traversed links, the amount of traffic to be transmitted and the modulation format to be adopted for transmission. Several alternative combinations of routes and modulation formats are considered and the classifier identifies the ones that will most likely satisfy the BER requirements. In \cite{Aladin2018}, a random forest classifier along with two other tools namely k-nearest neighbor and support vector machine are used. The authors in \cite{Aladin2018} use three of the above-mentioned classifiers to associate QoT labels with a large set of lightpaths to develop a knowledge base and find out which is the best classifier. It turns out from the analysis in \cite{Aladin2018}, that the support vector machine is better in performance than the other two but takes more computation time.

Two alternative approaches, namely \textit{network kriging}\footnote{Extensively used in the spatial statistics literature (see \cite{diggle2007springer} for details), kriging is closely related to Gaussian process regression (see chapter XV in \cite{murphy2014}).} (first described in \cite{chua2006network}) and \textit{norm $\mathcal{L}_2$ minimization} (typically used in network tomography \cite{castro2004network}), are applied in \cite{pointurier2011cross,sambo2010lightpath} in the context of QoT estimation: they rely on the installation of probe lightpaths that do not carry user data but are used to gather field measurements. The proposed inference methodologies exploit the spatial correlation between the QoT metrics of probes and data-carrying lightpaths sharing some physical links to provide an estimate of the Q-factor of already deployed or perspective lightpaths. These methods can be applied assuming either a centralized decisional tool or in a distributed fashion, where each node has only local knowledge of the network measurements. As installing probe lightpaths is costly and occupies spectral resources, the trade-off between number of probes and accuracy of the estimation is studied. Several heuristic algorithms for the placement of the probes are proposed in \cite{angelou2012optimized}. A further refinement of the methodologies which takes into account the presence of neighbor channels appears in \cite{sartzetakis2016quality}.

Additionally, a data-driven approach using a machine learning technique, Gaussian processes nonlinear regression (GPR), is proposed and experimentally demonstrated for performance prediction of WDM optical communication systems \cite{Thrane2017}. The core of the proposed approach (and indeed of any ML technique) is generalization: first the model is learned from the measured data acquired under one set of system configurations, and then the inferred model is applied to perform predictions for a new set of system configurations. The advantage of the approach is that complex system dynamics can be captured from measured data more easily than from simulations. Accurate BER predictions as a function of input power, transmission length, symbol rate and inter-channel spacing are reported using numerical simulations and proof-of-principle experimental validation  for a 24 $\times$ 28 GBd QPSK WDM optical transmission system.

Finally, a control and management architecture integrating an intelligent QoT estimator is proposed in \cite{bouda2017accurate} and its feasibility is demonstrated with implementation in a real testbed.

\subsection{Optical amplifiers control}
\begin{figure}
\centering
    \includegraphics[width=9cm]{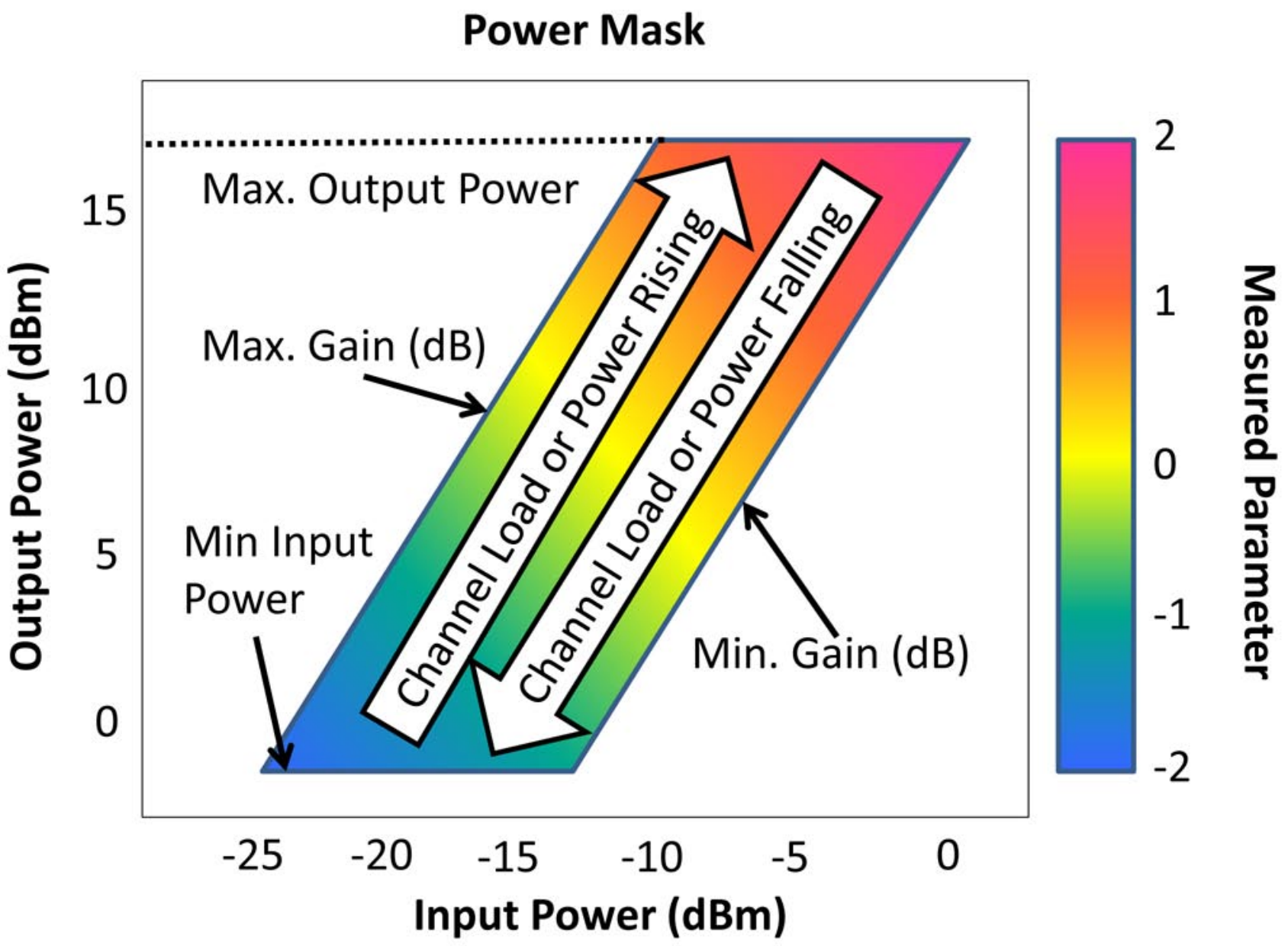}
    \caption{EDFA power mask \cite{bastos2013mapping}.} \label{fig:power_mask}
\end{figure}

The operating point of EDFAs influences their Noise Figure (NF) and gain flatness (GF), which have a considerable impact on the overall ligtpath QoT. The adaptive adjustment of the operating point based on the signal input power can be accomplished by means of ML algorithms. Most of the existing studies \cite{barboza2013self,bastos2013mapping,moura2016cognitive,oliveira2013demonstration,de2013edfa} rely on a preliminary amplifier  characterization process aimed at experimentally evaluating the value of the metrics of interest (e.g., NF, GF and gain control accuracy) within its power mask (i.e., the amplifier operating region, depicted in Fig. \ref{fig:power_mask}). 

The characterization results are then represented as a set of discrete values within the operation region. In EDFA implementations, state-of-the-art microcontrollers cannot easily obtain GF and NF values for points that were not measured during the characterization. Unfortunately, producing a large amount of fine grained measurements is time consuming. To address this issue, ML algorithms can be used to interpolate the mapping function over non-measured points.

For the interpolation, authors of \cite{barboza2013self,bastos2013mapping} adopt a NN implementing both feed-forward and backward error propagation. Experimental results with single and cascaded amplifiers report interpolation errors below 0.5 dB.
Conversely, a cognitive methodology is proposed in \cite{moura2016cognitive}, which is applied in dynamic network scenarios upon arrival of a new lightpath request: a knowledge database is maintained where measurements of the amplifier gains of already established lightpaths are stored, together with the lightpath characteristics (e.g., number of links, total length, etc.) and the OSNR value measured at the receiver. The database entries showing the highest similarities with the incoming lightpath request are retrieved, the vectors of gains associated to their respective amplifiers are considered and a new choice of gains is generated by perturbation of such values. Then, the OSNR value that would be obtained with the new vector of gains is estimated via simulation and stored in the database as a new entry. After this, the vector associated to the highest OSNR is used for tuning the amplifier gains when the new lightpath is deployed.

An implementation of real-time EDFA setpoint adjustment using the GMPLS control plane and interpolation rule based on a weighted Euclidean distance computation is described in \cite{oliveira2013demonstration} and extended in \cite{de2013edfa} to cascaded amplifiers.

Differently from the previous references, in \cite{huang2017dynamic} the issue of modelling the channel dependence of EDFA power excursion is approached by defining a regression problem, where the input feature set is an array of binary values indicating the occupation of each spectrum channel in a WDM grid and the predicted variable is the post-EDFA power discrepancy. Two learning approaches (i.e., the Ridge regression and Kernelized Bayesian regression models) are compared for a setup with 2 and 3 amplifier spans, in case of single-channel and superchannel add-drops. Based on the predicted values, suggestion on the spectrum allocation ensuring the least power discrepancy among channels can be provided.

\subsection{Modulation format recognition}
\begin{figure}
\centering
    \includegraphics[width=9cm]{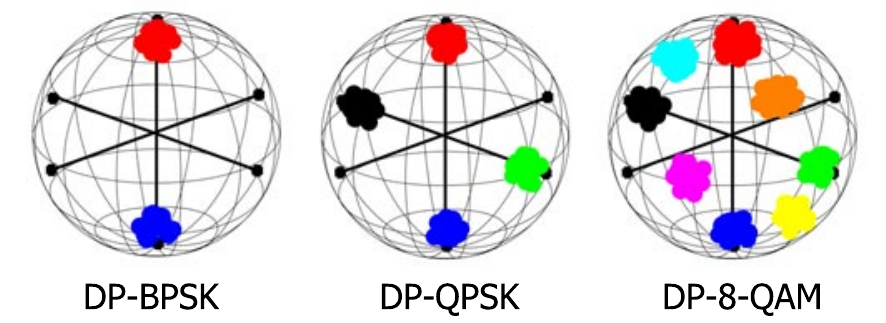}
    \caption{Stokes space representation of DP-BPSK, DP-QPSK and DP-8-QAM modulation formats \cite{borkowski2013stokes}.} \label{fig:stokes}
\end{figure}

The issue of autonomous modulation format identification in digital coherent receivers (i.e., without requiring information from the transmitter) has been addressed by means of a variety of ML algorithms, including k-means clustering \cite{gonzalez2010cognitive} and neural networks \cite{khan2012modulation,khan2016modulation}. 
Papers \cite{boada2015clustering} and \cite{borkowski2013stokes} take advantage of the Stokes space signal representation (see Fig. \ref{fig:stokes} for the representation of DP-BPSK, DP-QPSK and DP-8-QAM), which is not affected by frequency and phase offsets.

The first reference compares the performance of 6 unsupervised clustering algorithms to discriminate among 5 different formats (i.e. BPSK, QPSK, 8-PSK, 8-QAM, 16-QAM) in terms of True Positive Rate and running time depending on the OSNR at the receiver. For some of the considered algorithms, the issue of predetermining the number of clusters is solved by means of the \textit{silhouette coefficient}, which evaluates the tightness of different clustering structures by considering the inter- and intra-cluster distances.
The second reference adopts an unsupervised variational Bayesian expectation maximization algorithm to count the number of clusters in the Stokes space representation of the received signal and provides an input to a cost function used to identify the modulation format. The experimental validation is conducted over $k$-PSK (with $k=2,4,8$) and $n$-QAM (with $n=8,12,16$) modulated signals.

Conversely, features extracted from asynchronous amplitude histograms sampled from the eye-diagram after equalization in digital coherent transceivers
are used in \cite{khan2012modulation,khan2016modulation,zhang2016modulation} to train NNs. In \cite{khan2012modulation,khan2016modulation}, a NN is used for hierarchical extraction of the amplitude histograms' features, in order to obtain a compressed representation, aimed at reducing the number of neurons in the hidden layers with respect to the number of features. In \cite{zhang2016modulation}, a NN is combined with a genetic algorithm to improve the efficiency of the weight selection procedure during the training phase. Both studies provide numerical results over experimentally generated data: the former obtains 0\% error rate in discriminating among three modulation formats (PM-QPSK, 16-QAM and 64-QAM), the latter shows the tradeoff between error rate and number of histogram bins considering six different formats (NRZ-OOK, ODB, NRZ-DPSK, RZ-DQPSK, PM-RZ-QPSK and PM-NRZ-16-QAM).

\subsection{Nonlinearity mitigation}
One of the performance metrics commonly used for optical communication systems is the \textit{data-rate$\times$distance} product. Due to the fiber loss, optical amplification needs to be employed and, for increasing transmission distance, an increasing number of optical amplifiers must be employed  accordingly. Optical amplifiers add noise and to retain the signal-to-noise ratio optical signal power is increased. However, increasing the optical signal power beyond a certain value will enhance optical fiber nonlinearities which leads to Nonlinear Interference (NLI) noise. NLI will impact symbol detection and the focus of many papers, such as \cite{wang2016nonlinearity,wang2015nonlinear,Zibar:12,Zibar2016a,Zibar2016,shen2011fiber,giacoumidis2017reduction} has been on applying ML approaches to perform optimum symbol detection. 

In general, the task of the receiver is to perform optimum symbol detection. In the case when the noise has circularly symmetric Gaussian distribution, the optimum symbol detection is performed by minimizing the Euclidean distance between the received symbol $\mathbf{y}_k$ and all the possible symbols of the constellation alphabet, $s = {s_k|k = 1,...,M}$. This type of symbol detection will then have linear decision boundaries. For the case of memoryless nonlinearity, such as nonlinear phase noise, I/Q modulator and driving electronics nonlinearity, the noise associated with the symbol $\mathbf{y}_k$ may no longer be circularly symmetric. This means that the clusters in
constellation diagram become distorted (elliptically shaped instead of circularly symmetric in some cases). In those particular cases, optimum symbol detection is no longer based on Euclidean distance matrix, and the knowledge and full parametrization of the likelihood function, $p(\mathbf{y}_k|\mathbf{x}_k)$, is necessary. To determine and parameterize the likelihood function and finally perform optimum symbol detection, ML techniques, such as SVM, kernel density estimator, k-nearest neighbors and Gaussian mixture models can be employed. A gain of approximately 3 dB in the input power to the fiber has been achieved, by employing Gaussian mixture model in combination with expectation maximization, for 14 Gbaud DP 16-QAM transmission over a 800 km dispersion compensated link \cite{Zibar:12}.

Furthermore, in \cite{wang2016nonlinearity} a distance-weighted k-nearest neighbors classifier is adopted to compensate system impairments in zero-dispersion, dispersion managed and dispersion unmanaged links, with 16-QAM transmission, whereas in \cite{jarajreh2015artificial} NNs are proposed for nonlinear equalization in 16-QAM OFDM transmission (one neural network per subcarrier is adopted, with a number of neurons equal to the number of symbols). To reduce the computational complexity of the training phase, an Extreme Learning Machine (ELM) equalizer is proposed in \cite{shen2011fiber}. ELM is a NN where the weights minimizing the input-output mapping error can be computed by means of a generalized matrix inversion, without requiring any weight optimization step. 

SVMs are adopted in \cite{wang2015nonlinear,giacoumidis2017reduction}: in \cite{wang2015nonlinear}, a battery of $log_2(M)$ binary SVM classifiers is used to identify decision boundaries separating the points of a $M$-PSK constellation, whereas in \cite{giacoumidis2017reduction} fast Newton-based SVMs are employed to mitigate inter-subcarrier intermixing in 16-QAM OFDM transmission.

All the above mentioned approaches lead to a 0.5-3 dB improvement in terms of BER/Q-factor.

In the context of nonlinearity mitigation or in general, impairment mitigation, there are a group of references that implement equalization of the optical signal using a variety of ML algorithms like Gaussian mixture models \cite{Lu2018}, clustering \cite{XLu2018}, and artificial neural networks \cite{Li2018, Chuang2018, Pli2018, Jones2018, Hager2018, SLiu2018}. In \cite{Lu2018}, the authors propose a GMM to replace the soft/hard decoder module in a PAM-4 decoding process whereas in \cite{XLu2018}, the authors propose a scheme for pre-distortion using the ML clustering algorithm to decode the constellation points from a received constellation affected with nonlinear impairments. 

In references \cite{Li2018, Chuang2018, Pli2018, Jones2018, Hager2018, SLiu2018} that employ neural networks for equalization, usually a vector of sampled receive symbols act as the input to the neural networks with the output being equalized signal with reduced inter-symbol interference (ISI). In \cite{Li2018}, \cite{Chuang2018}, and \cite{Pli2018} for example, a convolutional neural network (CNN) would be used to classify different classes of a PAM signal using the received signal as input. The number of outputs of the CNN will depend on whether it is a PAM-4, 8, or 16 signal. The CNN-based equalizers reported in \cite{Li2018, Chuang2018, Pli2018} show very good BER performance with strong equalization capabilities. 

While \cite{Li2018, Chuang2018, Pli2018} report CNN-based equalizers, \cite{Hager2018} shows another interesting application of neural network in impairment mitigation of an optical signal. In \cite{Hager2018}, a neural network approximates very efficiently the function of digital back-propagation (DBP), which is a well-known technique to solve the non-linear Schroedinger equation using split-step Fourier method (SSFM) \cite{Ezra2008}. In \cite{Jones2018} too, a neural network is proposed to emulate the function of a receiver in a nonlinear frequency division multiplexing (NFDM) system. The proposed NN-based receiver in \cite{Jones2018} outperforms a receiver based on nonlinear Fourier transform (NFT) and a minimum-distance receiver. 

The authors in \cite{SLiu2018} propose a neural-network-based approach in nonlinearity mitigation/equalization in a radio-over-fiber application where the NN receives signal samples from different users in an Radio-over-Fiber system and returns a impairment-mitigated signal vector. 

An example of unsupervised k-means clustering technique applied on a received signal constellation to obtain a density-based spatial constellation clusters and their optimal centroids is reported in \cite{Zhang2018}. The proposed method proves to be an efficient, low-complexity equalization technique for a 64-QAM long-haul coherent optical communication system.
 
\subsection{Optical performance monitoring}

Artificial neural networks are well suited machine learning tools to perform optical performance monitoring as they can be used to learn the complex mapping between samples or extracted features from the symbols and optical fiber channel parameters, such as OSNR, PMD, Polarization-dependent loss (PDL), baud rate and CD. The features that are fed into the neural network can be derived using different approaches relying on feature extraction from: 1) the power eye diagrams (e.g., Q-factor, closure, variance, root-mean-square jitter and crossing amplitude, as in \cite{jargon2010optical,wu2009applications,khan2012optical,shen2010optical,Thrane2017,Zibar2016a}); 2) the two-dimensional eye-diagram and phase portrait \cite{anderson2009multi}; 3) asynchronous constellation diagrams (i.e., vector diagrams also including transitions between symbols \cite{jargon2010optical}); and 4) histograms of the asynchronously sampled signal amplitudes \cite{shen2010optical,khan2012optical}. The advantage of manually providing the  features to the algorithm is that the NN can be relatively simple, e.g., consisting of one hidden layer and up to 10 hidden units and does not require large amount of data to be trained. Another approach is to simply pass the samples at the symbol level and then use more layers that act as feature extractors (i.e., performing deep learning) \cite{tanimura2016osnr,Tanimura2018}. Note that this approach requires large amount of data due to the high dimensionality of the input vector to the NN. 

Besides the artificial neural network, other tools like Gaussian process models are also used which are shown to perform better in optical performance monitoring compared to linear-regression-based prediction models \cite{Meng2018}. The authors in \cite{Meng2018} also claims that sometimes simpler ML tools like the Gaussian Process (compared to ANN) can prove to be robust under noise uncertainties and can be easy to integrate into a network controller.


%% file: sections/Detailed_survey_of_ML_in_Networking_Problems.tex
\section{Detailed survey of machine learning in network layer domain}
\label{sec:5}
\begin{figure*}
\centering
    \includegraphics[width=0.9\textwidth]{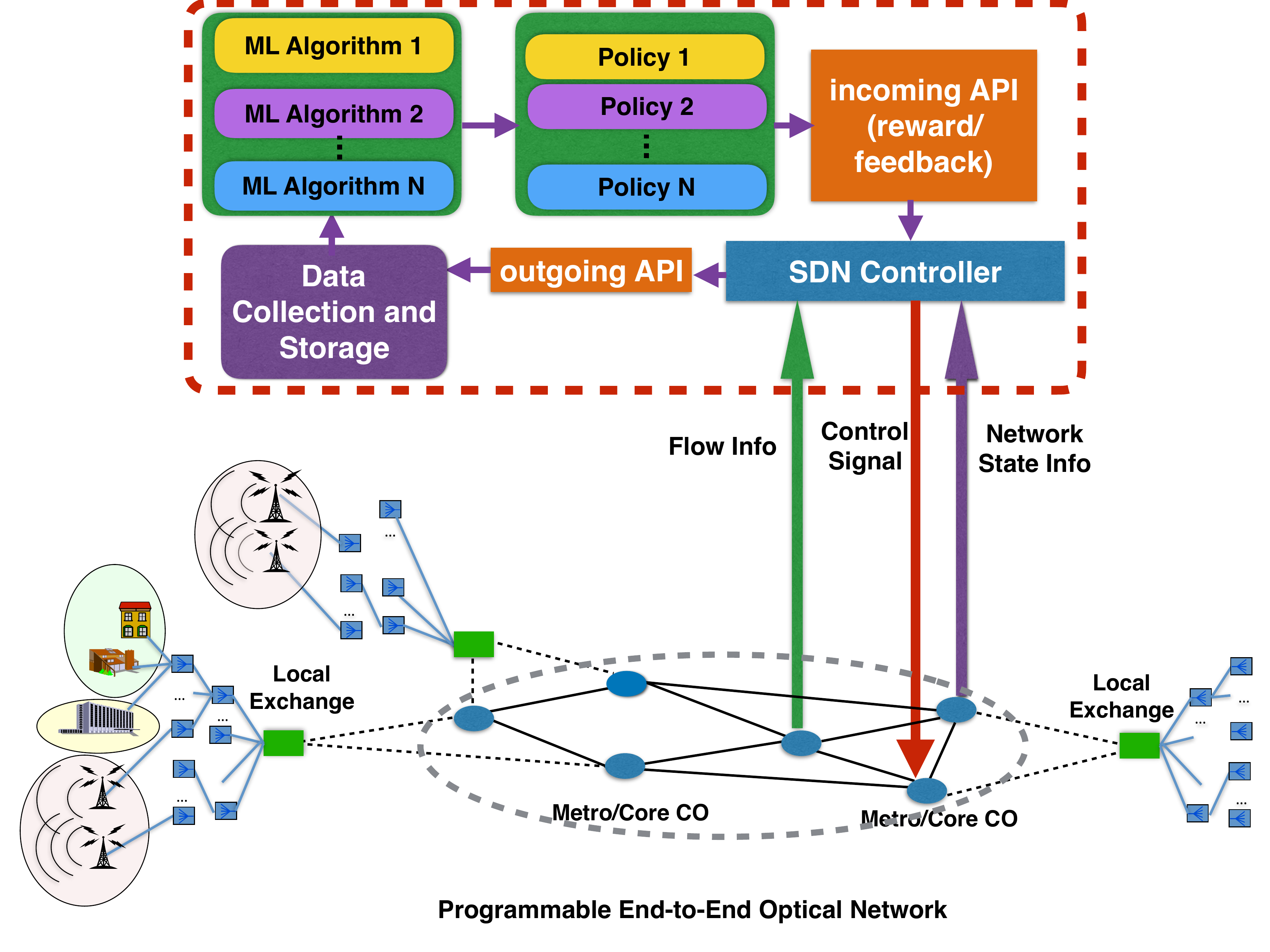}
    \caption{Schematic diagram illustrating the role of control plane housing the ML algorithms and policies for management of optical networks.} \label{fig:ML_net}
\end{figure*}



\subsection{Traffic prediction and virtual topology design}
Traffic prediction in optical networks is an important phase, especially in planning for resources and upgrading them optimally. Since one of the inherent philosophy of ML techniques is to learn a model from a set of data and `predict' the future behavior from the learned model, ML can be effectively applied for traffic prediction. 

For example, the authors in \cite{Fernandez2013}, \cite{Fernandez2015} propose Autoregressive Integrated Moving Average (ARIMA) method which is a supervised learning method applied on time series data \cite{ARIMA}. In both \cite{Fernandez2013} and \cite{Fernandez2015} the authors use ML algorithms to predict traffic for carrying out virtual topology reconfiguration. The authors propose a network planner and decision maker (NPDM) module for predicting traffic using ARIMA models. The NPDM then interacts with other modules to do virtual topology reconfiguration. 

Since, the virtual topology should adapt with the variations in traffic which varies with time, the input dataset in \cite{Fernandez2013} and \cite{Fernandez2015} are in the form of time-series data. More specifically, the inputs are the real-time traffic matrices observed over a window of time just prior to the current period. ARIMA is a forecasting technique that works very well with time series data \cite{ARIMA} and hence it becomes a preferred choice in applications like traffic predictions and virtual topology reconfigurations. Furthermore, the relatively low complexity of ARIMA is also preferable in applications where maintaining a lower operational expenditure as mentioned in \cite{Fernandez2013} and \cite{Fernandez2015}. 

In general, the choice of a ML algorithm is always governed by the trade-off between accuracy of learning and complexity. There is no exception to the above philosophy when it comes to the application of ML in optical networks. For example, in \cite{Morales2016} and \cite{Morales2017}, the authors present traffic prediction in an identical context as \cite{Fernandez2013} and \cite{Fernandez2015}, i.e., virtual topology reconfiguration, using NNs. 
A prediction module based on NNs is proposed which generates the source-destination traffic matrix. This predicted traffic matrix for the next period is then used by a decision maker module to assert whether the current virtual network topology (VNT) needs to be reconfigured. According to \cite{Morales2017}, the main motivation for using NNs is their better adaptability to changes in input traffic and also the accuracy of prediction 
of the output traffic based on the inputs (which are historical traffic). 

In \cite{Yu2018}, the authors propose a deep-learning-based traffic prediction and resource allocation algorithm for an intra-data-center network. The deep-learning-based model outperforms not only conventional resource allocation algorithms but also a single-layer NN-based algorithm in terms of blocking performance and resource occupation efficiency. The results in \cite{Yu2018} also bolsters the fact reflected in the previous paragraph about the choice of a ML algorithm. Obviously deep learning, which is more complex than a regular NN learning will be more efficient. Sometimes the application type also determines which particular variant of a general ML algorithm should be used. For example, recurrent neural networks (RNN), which best suits application that involve time series data is applied in \cite{WMo2018}, to predict baseband unit (BBU) pool traffic in a 5G cloud Radio Access Network. Since the traffic aggregated at different BBU pool comprises of different classes such as residential traffic, office traffic etc., with different time variations, the historical dataset for such traffic always have a time dimension. Therefore, the authors in \cite{WMo2018} propose and implement with good effect (a 7\% increase in network throughput and an 18\% processing resource reduction is reported) a RNN-based traffic prediction system. 

Reference \cite{Gifre2016} reports a cognitive network management module in relation to the Application-Based 
Network Operations (ABNO) framework, with specific focus on ML-based traffic prediction for VNT reconfiguration. However, \cite{Gifre2016} does not mention about the details of any specific ML algorithm used for the purpose of VNT reconfiguration. On similar lines, \cite{Ohba2017} proposes bayesian inference to estimate network traffic and decide whether to reconfigure a given virtual network.

While most of the literature focuses on traffic prediction using ML algorithms with a specific view of virtual network topology reconfigurations, \cite{Troia2017} presents a general framework of traffic pattern estimation from call data records (CDR). \cite{Troia2017} uses real datasets from service providers and operates matrix factorization and clustering based algorithms to draw useful insights from those data sets, which can be utilized to better engineer the network resources. More specifically, \cite{Troia2017} uses CDRs from different base stations from the city of Milan. The dataset contains information like cell ID, time interval of calls, country code, received SMS, sent SMS, received calls, sent calls, etc., in the form of a matrix called CDR matrix. Apart from the CDR matrix, the input dataset also includes a point-of-interest (POI) matrix which contains information about different points of interests or regions most likely visited corresponding to each base station. All these input matrices are then applied to a ML clustering algorithm called non-negative matrix factorization (NMF) and a variant of it called collective NMF (C-NMF). The output of the algorithms factors the input matrices into two non-negative matrices one of which gives the different types basic traffic patterns and the other gives similarities between base stations in terms of the traffic patterns. 

While many of the references in the literature focus on one or few specific features when developing ML algorithms for traffic prediction and virtual topology (re)configurations, others just mention a general framework with some form of `cognition' incorporated in association with regular optimization algorithms. For example, \cite{Fernandez2012Energy} and \cite{Fernandez2012Surv} describes a multi-objective Genetic Algorithm (GA) for virtual topology design. No specific machine learning algorithm is mentioned in \cite{Fernandez2012Energy} and \cite{Fernandez2012Surv}, but they adopt adaptive fitness function update for GA. Here they use the principles of reinforcement learning where previous solutions of the GA for virtual topology design are used to update the fitness function for the future solutions. 

\subsection{Failure management}
ML techniques can be adopted to either identify the exact location of a failure or malfunction within the network or even to infer the specific type of failure. 
In \cite{christodoulopoulos2016exploiting}, network kriging is exploited to localize the exact position of failure along network links, under the assumption that the only information available at the receiving nodes (which work as monitoring nodes) of already established lightpaths is the number of failures encountered along the lightpath route. If unambiguous localization cannot be achieved, lightpath probing may be operated in order to provide additional information, which increases the rank of the routing matrix. Depending on the network load, the number of monitoring nodes necessary to ensure unambiguous localization is evaluated.
Similarly, in \cite{ruiz2016service} the measured time series of BER and received power at lightpath end nodes are provided as input to a Bayesian network which individuates whether a failure is occurring along the lightpath and try to identify the cause (e.g., tight filtering or channel interference), based on specific attributes of the measurement patterns (such as maximum, average and minimum values, presence and amplitude of steps). The effectiveness of the Bayesian classifier is assessed in an experimental testbed: results show that only 0.8\% of the tested instances were misclassified.

Other instances of application of Bayesian models to detect and diagnose failures in optical networks, especially GPON/FTTH, are reported in \cite{Tembo2016} and \cite{Gosselin2017}. In \cite{Tembo2016}, the GPON/FTTH network is modeled as a Bayesian Network using a layered approach identical to one of their previous works \cite{Tembo2015}. The layer 1 in this case actually corresponds to the physical network topology consisting of ONTs, ONUs and fibers. Failure propagation, between different network components depicted by layer-1 nodes, is modeled in layer 2 using a set of directed acyclic graphs interconnected via the layer 1. The uncertainties of failure propagation are then handled by quantifying strengths of dependencies between layer 2 nodes with conditional probability distributions estimated from network generated data. 
However, some of these network generated data can be missing because of improper measurements or non-reporting of data. 
An Expectation Maximization (EM) algorithm is therefore used to handle missing data for root-cause analysis of network failures and helps in self-diagnosis. Basically, the EM algorithm estimates the missing data such that the estimate maximizes the expected log-likelihood function based on a given set of parameters. In \cite{Gosselin2017} a similar combination of Bayesian probabilistic models and EM is used for failure diagnosis in GPON/FTTH networks.

In the context of failure detection, in addition to Bayesian networks, other machine learning algorithms and concepts have also been used. For example, in \cite{Vela_JLT_failures}, two ML based algorithms are described based on regression, classification, and anomaly detection. The authors propose a BER anomaly detection algorithm which takes as input historical information like maximum BER, threshold BER at set-up, and monitored BER per lightpath and detects any abrupt changes in BER which might be a result of some failures of components along a lightpath. This BER anomaly detection algorithm, which is termed as BANDO, runs on each node of the network. The outputs of BANDO are different events denoting whether the BER is above a certain threshold or below it or within a pre-defined boundary. 

This information is then passed on to the input of another ML based algorithm which the authors term as LUCIDA. LUCIDA runs in the network controller and takes historic BER, historic received power, and the outputs of BANDO as input. These inputs are converted into three features that can be quantified by time series and they are as follows: 1) Received power above the reference level (PRXhigh); 2) BER positive trend (BERTrend); and 3) BER periodicity (BERPeriod). LUCIDA computes these features' probabilities and the probabilities of possible failure classes and finally maps these feature probabilities to failure probabilities. In this way, LUCIDA detects the most likely failure cause from a set of failure classes. 

Another notable use case for failure detection in optical networks using ML concepts appear in \cite{Vela:18}. Two algorithms are proposed viz., Testing optIcal Switching at connection SetUp time (TISSUE) and FailurE causE Localization for optIcal NetworkinG (FEELING). The TISSUE algorithm takes the values of estimated BER calculated at each node across a lightpath and the measured BER and compares them. If the differences between the slopes of the estimated and theoretical BER is above a certain threshold a failure is anticipated. While it is not clear from \cite{Vela:18} whether the estimation of BER in the TISSUE algorithm is based on ML methods, the FEELING algorithm applies two very well-known ML methods viz., decision tree and SVM. 

In FEELING, the first step is to process the input dataset in the form of ordered pairs of frequency and power for each optical signal and transform them into a set of features. The features include some primary features like the power levels across the central frequency of the signal and also the power around other cut-off points of the signal spectrum (interested readers are encouraged to look into \cite{Vela:18} for further details). In context of the FEELING algorithm, some secondary features are also defined in \cite{Vela:18} which are linear combinations of the primary features. The feature-extraction process is undertaken by a module named FeX. The next step is to input these features into a multi-class classifier in the form of a decision tree which outputs a predicted class among three options: `Normal’, `LaserDrift’ and `FilterFailure’; and ii) a subset of relevant signal points for the predicted class. Basically, the decision tree contains a number of decision rules to map specific combinations of feature values to classes. This decision-tree-based component runs in another module named signal spectrum verification (SSV) module. The FeX and SSV modules are located in the network nodes. There are two more modules called signal spectrum comparison (SSC) module and laser drift estimator (LDE) module which runs on the network controller. 

In the SSC module, a similar classification process takes place as in SSV. But here a signal is diagnosed based on the different classes of failures just due to filtering. Here the three classes are:  Normal, FilterShift and TightFiltering. The SSC module uses Support Vector Machines to classify the signals based on the above three classes. First, the SVM classifies whether the signal is `Normal' or has suffered a filter-related failure. Next, the SVM classifies the signal suffering from filter-related failures into two classes based on whether the failure is due to tight filtering or due to filter shift. Once these classifications are done, the magnitude of failures related to each of these classes are estimated using some linear regression based estimator modules for each of the failure classes. Finally, all these information provided by the different modules described so far, are used in the FEELING algorithm to return a final list of failures. 

\begin{figure}
\centering
    \includegraphics[width=8.7cm]{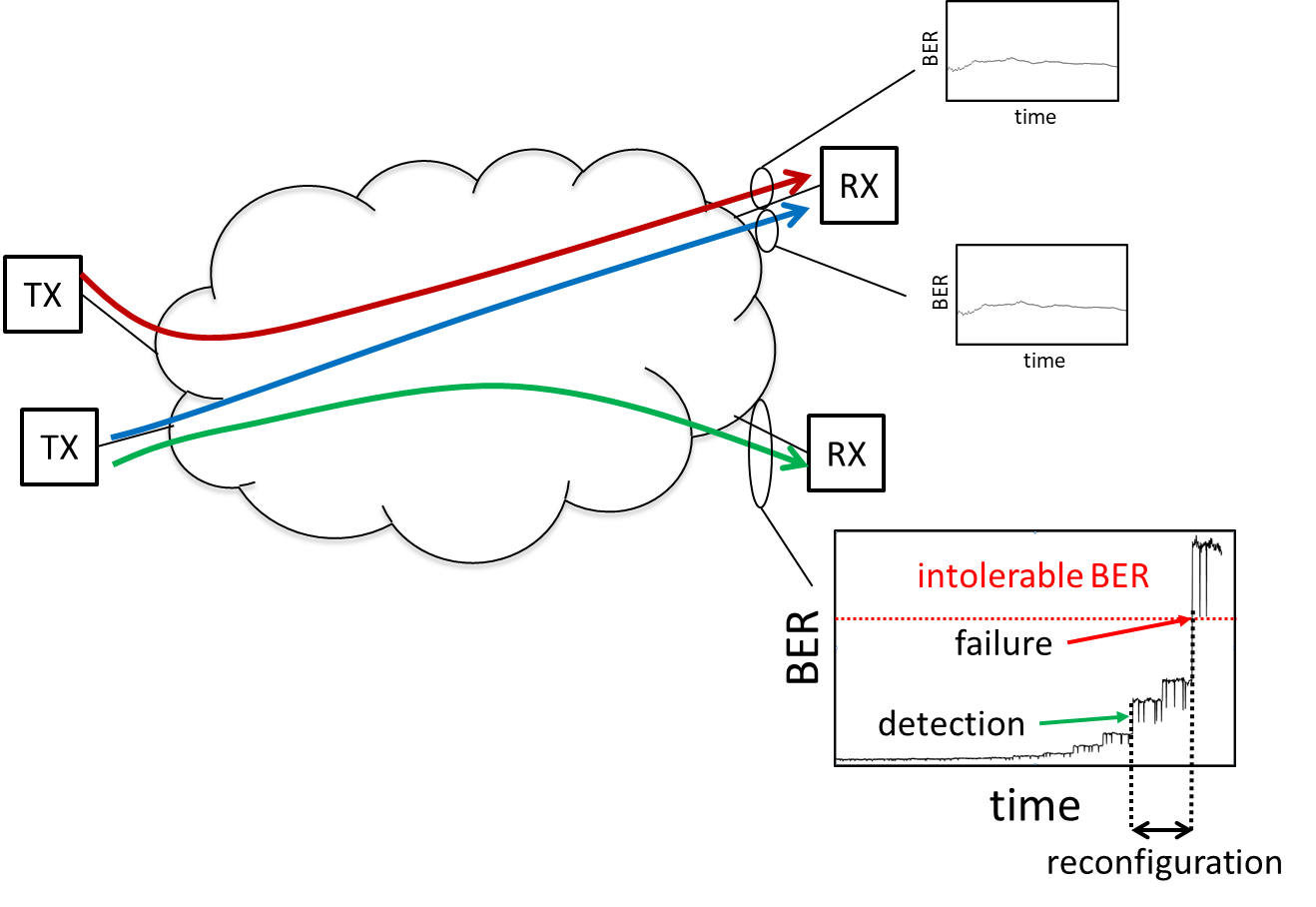}
    \caption{The failure detection and identification framework adopted in \cite{Shahkarami2018}.} \label{fig:detection}
\end{figure}

A similar multi-ML algorithm based framework like \cite{Vela:18} for failure detection and classification is also proposed in \cite{Shahkarami2018} and \cite{Rafique2018}. In \cite{Shahkarami2018} several ML algorithms are used and, by tuning several model parameters, such as BER sampling time and amount of BER data needed to train the models, one or more proper optimized algorithm(s) is/are chosen from Binary and Multiclass SVMs, Random Forests and neural networks. Moreover, in paper \cite{Shahkarami2018}, the authors propose the detection and cause identification algorithms suggesting that a network operator, able to early-detect a failure (and identify its cause) before a critical BER threshold is reached, can proactively re-route the affected traffic onto a new lightpath, so as to minimize SLA violation and enhance (i.e., speed up) failure recovery procedures (see Fig. \ref{fig:detection}).

In \cite{Rafique2018}, optical power levels, amplifier gain, shelf temperature, current draw, internal optical power are used to predict failures using statistical regression and neural network based algorithms that sit in the SDN controllers. 

\subsection{Flow classification}
Another popular area of ML application for optical networks is flow classification.
In \cite{Jayaraj2008OBS_TCP} for example, a framework is described that observes different types of packet loss in optical burst-switched (OBS) networks. It then classifies the packet loss data as congestion loss or contention loss using a Hidden Markov Model (HMM) and EM algorithms. 

Another example of flow classification is presented in \cite{Viljoen2016}. Here a NN is trained to classify flows in an optical data center network. The feature vector includes a 5-tuple (source IP address, destination IP address, source port, destination port, transport layer protocol). Packet sizes and a set of intra-flow timings within the first 40 packets of a flow, which roughly corresponds to the first 30 TCP segments, are also used as inputs to improve the training speed and to mitigate the problem of `disappearing gradients' while using gradient descent for back-propagation. 

The main outcome of the NN used in \cite{Viljoen2016} is the classification of mice and elephant flows in the data center (DC). The type of neural network used is a multi-layer perceptron (MLP) with four hidden layers as MLPs are relatively simpler to implement. The authors of \cite{Viljoen2016} also mention the high levels of true negative classification associated with MLPs, and comment on importance of ensuring that mice do not flood the optical interconnections in the DC network. In general, mice flows do actually outnumber elephant flows in a practical DC network, and therefore the authors in \cite{Viljoen2016} suggest to overcome this class imbalance between mice and elephant flows by training the NN with a non-proportional amount of mice and elephant flows. 

\subsection{Path computation}
Path computation or selection, based on different physical and network layer parameters, is a commonly studied problem in optical networks. In Section \ref{sec:4} for example, physical layer parameters like QoT, modulation format, OSNR, etc. are estimated using ML techniques. The main aim is to make a decision about the best optical path to be selected among different alternatives. The overall path computation process can therefore be viewed as a cross-layer method with application of machine learning techniques in multiple layers. In this subsection we identify references \cite{Kiran2007OBS} and \cite{Tronco2016} that addresses the path computation/selection in optical networks from a network layer perspective. 

In \cite{Kiran2007OBS} the authors propose a path and wavelength selection strategy for OBS networks to minimize burst-loss probability. The problem is formulated as a multi-arm bandit problem (MABP) and solved using Q-learning. An MABP problem comes from the context of gambling where a player tries to pull one of the arms of a slot machine with the objective to maximize sum of rewards over many such pulls of arms. In the OBS network scenario, the authors in \cite{Kiran2007OBS} use the concept of path selection for each source-destination pair as pulling of one of the arms in a slot machine with the reward being minimization of burst-loss probability. In general the MABP problem is a classical problem in reinforcement learning and the authors propose Q-learning to solve this problem because other methods does not scale well for complex problems. Furthermore, other methods of solving MABP, like dynamic programming, Gittins indices, and learning automata prove to be difficult when the reward distributions (i.e., the distributions of the burst-loss probability in case of the OBS scenario) are unknown. The authors in \cite{Kiran2007OBS} also argue that the Q-learning algorithm has a guaranteed convergence compared to other methods of solving the MABP problem.

In \cite{Tronco2016} a control plane decision making module for QoS-aware path computation is proposed using a Fuzzy C-Means Clustering (FCM) algorithm. The FCM algorithm is added to the software-defined optical network (SDON) control plane in order to achieve better network performance, when compared with a non-cognitive control plane. The FCM algorithm takes traffic requests, lightpath lengths, set of modulation formats, OSNR, BER etc., as input and then classifies each lightpath with the best possible parameters of the physical layer. The output of the classification is a mapping of each lightpath with a different physical layer parameter and how closely a lightpath is associated with a physical layer parameter in terms of a membership score. This membership score information is then utilized to generate some rules based on which real time decisions are taken to set up the lightpaths. 

As we can see from the overall discussion in this section, different ML algorithms and policies can be used based on the use cases and applications of interest. Therefore, one can envisage a concise control plane for the next generation optical networks with a repository of different ML algorithms and policies as shown in Fig. \ref{fig:ML_net}. The envisaged control plane in Fig. \ref{fig:ML_net} can be thought of as the `brain' of the network that interacts constantly with the `network body' (i.e., different components like transponders, amplifies, links etc.) and react to the `stimuli' (i.e., data generated by the network) and perform certain `actions' (i.e., path computation, virtual topology (re)configurations, flow classification etc.). A setup on similar lines as discussed above is presented in \cite{Sgambelluri2018} where a ML-based monitoring module drives a Reconfigurable Optical Add/Drop Multiplexer (ROADM) controller which is again controlled by a SDN controller. 

%% file: sections/Comparison.tex
\section{Evaluation of machine learning algorithms in optical networks}\label{sec:comparison}

In this section, we provide a more quantitative comparison of some of the ML applications described in Section \ref{sec:3}.
To do this, we first provide an overview of the typical performance metrics adopted in ML.
Then, we select some of the studies discussed in Sections \ref{sec:4} and \ref{sec:5}, and we concentrate on how the ML algorithms used these papers are quantitatively compared using these performance metrics. For each paper, we also provide a quick description of the main outcome of this comparison.

\subsection{Performance metrics}

\begin{table*}[t]
\caption{Comparison of ML algorithms and performance metrics for a selection of existing papers.}
\centering
\begin{tabular}
{p{2.5cm} c p{3.2cm} p{2.5cm} p{6.7cm}}
\toprule
Use Case	&Ref.	&Adopted algorithms	&Metrics	&Outcome\\
\midrule
\midrule
QoT estimation (BER classification)	&\cite{de2013cognitive}	&Naive Bayes, Decision tree, RF, J4.8 tree, CBR	&Accuracy, false positives	&CBR has highest accuracy (above 99\%) with low false positive (0.43\%), decision tree reaches lowest false positive (0.02\%) at the price of much lower accuracy (86\%)\\
\midrule
QoT estimation (BER classification)	&\cite{barletta17QoT}	&KNN, RF	&Accuracy, AUC, running time	&RF has higher AUC and accuracy than KNN, the training time of RF is higher than KNN but the testing time is at least one order of magnitude lower than RNN\\
\midrule
QoT estimation (BER classification)	&\cite{Aladin2018}	&KNN, RF, SVM	&Accuracy, Confusion Matrix, ROC curves	&SVM has the best accuracy among all three ML algorithms, accuracy improves with size of Knowledge Base (KB)\\
\midrule
MF recognition in Stokes space	&\cite{boada2015clustering} &K-means, EM, DBSCAN, OPTICS, spectral clustering, Maximum-likelihood	&Running time, minimum OSNR to achieve 95\% accuracy	&Maximum likelihood requires lowest OSNR level and has very low running time (comparable to OPTICS, which has lowest running time but requires much higher OSNR level)\\
\midrule
Failure Management	&\cite{Tembo2016}, \cite{Gosselin2017} &Bayesian Inference, EM	&Confusion Matrix	&The failure detection based on learning of the network parameters is more accurate compared to the case where an expert sets the parameters based on certain deterministic rules\\
\midrule
Failure Management	&\cite{Shahkarami2018} &NN, RF, SVM	&Accuracy versus model parameters (BER sampling time, amount of BER data etc.)	&With right model parameters, binary SVM can reach up to 100\% accuracy for failure detection\\
\midrule
Flow Classification (Loss classification in OBS networks)	&\cite{Jayaraj2008OBS_TCP}	&HMM, EM	&Misclassification probability (similar to FPR)	&HMM has better accuracy and has lower misclassification probability for static traffic type compared to dynamic traffic, the misclassification probability also goes down with increasing number of wavelengths per link\\
\bottomrule
\end{tabular}
\label{tab:comparison}
\end{table*}

When applying ML to a classification problem, a common approach to evaluate the ML-algorithm performance is to show its classification accuracy and a meadure of the algorithm complexity, usually expressed in the form of training-phase duration. 
Classification accuracy represents the fraction of the test samples which are correctly classified. Although this metric is intuitive, it turns out to be a poor metric in complex classification problems, especially when the available dataset contains an amount of samples largely unbalanced among the various classes (e.g., a binary dataset where 90\% of samples belongs to one class). In these cases, the following and other measures can be used: 
\begin{itemize}
\item \textit{Confusion matrix}: Given a binary classification problem, where samples in the test set belong to either a \textit{positive} or a \textit{negative} class, the confusion matrix gives a complete overview of the classifier performance, showing 1) the true positives ($TP$) and true negatives ($TN$), i.e., the number of samples of the true and false class, respectively, which have been correctly classified, and 2) the false positives ($FP$) and false negatives ($FN$), i.e., the number of samples of the true and false class, respectively, which have been misclassified. Note that, using these definitions, accuracy can be expressed as $(TP+TN)/(TP+TN+FP+FN)$.
\item \textit{True Positive Rate}, $TPR=TP/(TP+FN)$: This metric falls in the $[0,1]$ range and captures the ability of identifying actually positive samples in the test set (i.e., the larger, the better).
\item \textit{False Positive Rate}, $FPR=FP/(FP+TN)$: Also this metric falls in the $[0,1]$ range, and it represents the fraction of negative samples in the test set that are incorrectly classified as positive (i.e., the lower, the better).
\item \textit{Receiver operating characteristic (ROC) curve}: In a binary classifier, an arbitrary threshold $\gamma$ can be set to distinguish between true and false instances; by increasing the value of $\gamma$, we reduce the number of instances that we classify as positive and increase the number of samples that we classify as negative; this has the effect of decreasing $TP$ while correspondingly increasing $FN$, and increasing $TN$ while correspondingly decreasing $FP$; hence, both the TPR and the FPR are reduced. For different values of $\gamma$, the ROC curve plots the $TPR$ (on the vertical axis) against the $FPR$ (on the horizontal axis). For $\gamma=1$, all samples are classified as negative, therefore $TPR=FPR=0$. Conversely, for $\gamma=0$, all samples are classified as positive, hence $TPR=FPR=1$.
For any classifier, its ROC curve always connects these two extremes. Classifiers capturing useful information yield a ROC curve above the diagonal in the $(FPR,TPR)$ plane, and aim at approaching the ideal classifier, which interconnects points (0,0), (0,1) and (1,1).
\item \textit{Area under the ROC curve (AUC)}: The AUC takes values in the $[0,1]$ range and captures how much a given classifier approaches the performance of an ideal classifier. While the ROC curve is an efficient graphical means to evaluate the performance of a classifier, the AUC is a synthetic numerical measure to indicate algorithm performance independently from the specific choice of the threshold $\gamma$. 
\item \textit{Akaike Information Criteria (AIC)}: This is a metric that captures the goodness of fit for a particular model. It measures the deviation of a chosen statistical model from the `true model' by defining a criteria which is a mathematical function of the number of estimated parameters by the model and the maximum likelihood function. The model with minimum AIC is considered as the best model to fit a given dataset \cite{Akaike2011}. 
\item Metrics from the optical networking field: Besides numerical and graphical metrics traditionally used in the ML context, measures from the networking field can be also adopted in combination with such metrics, in order to have a quantitative understanding of how the ML algorithm impacts on the optical network/system. E.g., an operator might be interested in the minimum number of optical performance monitors to deploy along a lightpath to correctly classify a degraded transmission with a given accuracy; similarly, the minimum OSNR and/or signal power level required at an optical receiver to correctly recognize the adopted MF. 
Furthermore, an operator might also wonder how often BER samples should be collected to predict or correctly localize an optical failure along a lightpath with a certain accuracy.
\end{itemize}

\subsection{Quantitative algorithms comparison}

We now provide a schematic comparison of some ML algorithms focusing on some of the use cases discussed in Section \ref{sec:3}. To perform this comparison, we select, among the papers surveyed in Sections \ref{sec:4} and \ref{sec:5}, those where different ML algorithms have been applied and compared with a same data set. Note that a fair quantitative comparison between algorithms in different papers is hard due to the fact that, in each paper, the various algorithms have been designed to fit with the specific available data set. As a consequence, a given algorithm may perform incredibly well if applied to a certain data set, but at the same time it may exhibit poor performance if the data set is changed, though not substantially.

Table \ref{tab:comparison} provides such overview, highlighting, for each considered use case and corresponding reference, the ML algorithms and the evaluation metrics used for the comparison. In the table we also  provide a synthetic description of the paper outcome.

%% file: sections/Discussion_and_Future_Directions.tex
\section{Discussion and future directions}
\label{sec:6}
In this section we discuss our vision on how this research area will expand in next years, focusing on some specific areas that we believe will require more attention during the next years.

\textit{\textbf{ML methodologies.}} We notice how the vast majority of existing studies adopting ML in optical networks use offline supervised learning methods, i.e., assume that the ML algorithms are trained with historical data before being used to take decisions on the field. This assumption is often unrealistic for optical communication networks, where scenarios dynamically evolve with time due, e.g., to traffic variations or to changes in the behavior of optical components caused by aging.
We thus envisage that, after learning from a batch of available past samples, other types of algorithms, in the field of semi-supervised and/or unsupervised ML, could be implemented to gradually take in novel input data as they are made available by the network control plane. Under a different perspective, re-training of supervised mechanisms must be investigated to extend their applicability to, e.g., different network infrastructures (the training on a given topology might not be valid for a different topology) or to the same network infrastructure at a different point in time (the training performed in a certain week/month/year might not be valid anymore after some time).
In a more general sense, novel ML techniques, developed ad-hoc for optical-networking  problems might emerge. Consider, e.g., active ML algorithms, which can interactively ask the user to observe training data with specific characteristics. This way, the number of samples needed to build an accurate prediction model can be consistently
reduced, which may lead to significant savings in case the dataset generation process is costly (e.g., when probe lightpaths have to be deployed).

\textit{\textbf{Data availability.}} As of today, vendors and operators have not yet disclosed large set of field data to test the practicality of existing solutions. This problem might be partially addressed by emulating relevant events, as failures or signal degradations, over optical-network testbeds, even though it is simply impossible to reproduce the diversity of scenarios of a real network in a lab environment. Moreover, even in situations of complete access  to real data, for some of the use cases mentioned before, in practical assets it is difficult to collect extensive datasets during faulty operational conditions, since networks are typically dimensioned and managed via conservative design approaches which make the probability of failures negligible (at the price of under-utilization of network resources). 

\textit{\textbf{Timescales.}} Scarce attention has so far been devoted to the fact that different applications might have very different timescales over which monitored data show observable and useful pattern changes (e.g., aging would make component behaviour vary slowly over time, while traffic varies quickly, and at different timescales, e.g., burst, daily, weekly, yearly level. Understading the right timescale for the monitoring of the parameters to be fed into ML algorithms is not only important to optimize the accuracy of the algorithm (and hence system performance), but it is fundamental to dimension the amout of control/monitoring bandwidth needed to actually implement the ML-based system. If a ML-algorithm works perfectly, but it requires a huge amount of data to be sampled extremely frequently, then the additional control bandwidth required will hinder the practical application of the algorithm.

\textit{\textbf{A complete cognitive control system.}} Another important consideration is that all existing ML-based solutions have addressed specific and isolated issues in optical communications and networking. Considering that software defined networking has been demonstrated to be capable of successfully converging control through multiple network layers and technologies, such a unified control could also coordinate (orchestrate) several different applications of ML, to provide a holistic design for flexible optical networks.
In fact, as seen in the literature, ML algorithms can be adopted to estimate different system characteristics at different layers, such as QoT, failure occurrences, traffic patterns, etc., some of which are mutually dependent (e.g., the QoT of a lightpath is highly related to the presence of failures along its links or in the traversed nodes), whereas others do not exhibit dependency (e.g., traffic patterns and fluctuations typically do not show any dependency on the status of the transmission equipment). More research is needed to explore the applicability and assess the benefits of ML-based unified control frameworks where all the estimated variables can be taken into account when making decisions such as where to route a new lightpath (e.g., in terms of spectrum assignment and core/mode assignment), when to re-route an existing one, or when to modify transmission parameters such as modulation format and baud rate. 

\textit{\textbf{Failure recovery.}} Another promising and innovative area for ML application paired with SDN control is network failure recovery. State-of-the-art optical network control tools are tipically configured as rule-based expert systems, i.e., a set of expert rules (IF $<$conditions$>$ THEN $<$actions$>$) covering typical failure scenarios. Such rules are specialized and deterministic and usually in the order of a few tens, and cannot cover all the possible cases of malfunctions. The application of ML to this issue, in addition to its ability to take into account relevant data across all the layers of a network, could also bring in probabilistic characterization (e.g., making use of Gaussian processes, output probability distributions rather than single numerical/categorical values) thus providing much richer information with respect to currently adopted threshold-based models.

\textit{\textbf{Visualization}}. Developing effective visualization tools to make the information-rich outputs produced by ML algorithms immediately accessible and comprehensible to the end users is a key enabler for seamless integration of ML techniques in optical network management frameworks. Though some preliminary research steps in such direction have been done (see, e.g., \cite{vela2018applying}, where bubble charts and spectrum color maps are employed to visualize network links experiencing high BER), design guidelines for intuitive visualization approaches depending on the specific aim of ML usage (e.g., network monitoring, failure identification and localization, etc.) have yet to be investigated and devised.

\textit{\textbf{Commercialization and standardization.}} Though in its infancy, applications of ML to optical networking have already attracted the interest of network operators and optical equipment vendors, and it is expected that this attention will grow rapidly in the near future. Among the others, we notice some activities on QoT estimation optimization for margin reduction and error-aware rerouting \cite{SimsarianNAOG2017}, on low-margin optical network design \cite{Coriant2018}, on traffic prediction \cite{ChoudhuryJOCN18} and anomaly detection \cite{CoteJOCN18}.
Furthermore, also standardization bodies have started looking at the application of ML for the resolution of networking problems. Although, to the best of our knowledge, no specific activity is currently undergoing with dedicated focus on optical networks,
it is worth mentioning, e.g., ITU-T focus group on ML \cite{ITUML5G}, whose activities are concentrated on various aspects of future networking, such as architectures, interfaces, protocols, algorithms and data formats.

\textit{\textbf{Optics for Machine Learning (vs. Machine Learning for optics)}}. Finally, an interesting, though speculative, area of future research is the application of ML to all-optical devices and networks. Due to their inherent non-linear behaviour, optical components could be interconnected to form structures capable of implementing learning tasks \cite{woods2012optical}. This approach represents an all-optical alternative to traditional software implementations. In \cite{brunner2013high}, for example, semiconductor laser diodes were used to create a photonic neural network via time-multiplexing, taking advantage of their nonlinear reaction to power injection due to the coupling of amplitude and phase of the optical field. In \cite{fiers2012optical}, a ML method called \lq\lq reservoir computing'' is implemented via a nanophotonic reservoir constituted by a network of coupled crystal cavities. Thanks to their resonating behavior, power is stored in the cavities and generates nonlinear effects. The network is trained to reproduce periodic patterns (e.g., sums of sine waves).

To conclude, the application of ML to optical networking is a fast-growing research topic, which sees an increasingly strong participation from industry and academic researchers. While in this section we could only provide a short discussion on possible future directions, we envisage that many more research topic will soon emerge in this area.

%% file: sections/Conclusion.tex
\section{Conclusion}
\label{sec:conc}
Over the past decade, optical networks have been growing `smart' with the introduction of software defined networking, coherent transmission, flexible grid, to name only few arising technical and technological directions. 
The combined progress towards high-performance hardware and intelligent software, integrated through an SDN platform provides a solid base for promising innovations in optical networking. 
Advanced machine learning algorithms can make use of the large quantity of data available from network monitoring elements to make them `learn' from experience and make the networks more agile and adaptive. 

Researchers have already started exploring the application of machine learning algorithms to enable smart optical networks and in this paper we have summarized some of the work carried out in the literature and provided insight into new potential research directions.


%% file: sections/glossary.tex
\section*{Glossary}\label{sec:glossary}


\begin{supertabular}{l l}
\textbf{ABNO}	&Application-Based Network Operations \\
\textbf{ANN}	&Artificial Neural Network\\
\textbf{API}	&Application Programming Interface\\
\textbf{AUC}	&Area Under the ROC Curve\\
\textbf{ARIMA}	&Autoregressive Integrated Moving Average\\
\textbf{BBU} 	&Baseband Unit\\
\textbf{BER} 	&Bit Error Rate\\
\textbf{BPSK}	&Binary Phase Shift Keying\\
\textbf{BVT} 	&Bandwidth Variable Transponders\\
\textbf{CBR} 	&Case Based Reasoning\\
\textbf{CD} 	&Chromatic Dispersion\\
\textbf{CDR} 	&Call Data Records\\
\textbf{CNN}	&Convolutional Neural Network\\
\textbf{C-NMF}	&Collective Non-negative Matrix Factorization\\
\textbf{CO}		&Central Office\\
\textbf{DBP} 	&Digital Back Propagation\\
\textbf{DC} 	&Data Center\\
\textbf{DP} 	&Dual Polarization\\
\textbf{DQPSK}	&Differential Quadrature Phase Shift Keying\\
\textbf{EDFA}	&Erbium Doped Fiber Amplifier\\
\textbf{ELM}	&Extreme Learning Machine\\
\textbf{EM}		&Expectation Maximization\\
\textbf{EON}	&Elastic Optical Network\\
\textbf{FCM}	&Fuzzy C-Means Clustering\\
\textbf{FN}		&False negatives\\
\textbf{FP}		&False positives\\
\textbf{FTTH}	&Fiber-to-the-home\\
\textbf{GA}		&Genetic Algorithm\\
\textbf{GF}		&Gain flatness\\
\textbf{GMM}	&Gaussian Mixture Model\\
\textbf{GMPLS}	&Generalized Multi-Protocol Label Switching\\
\textbf{GPON}	&Gigabit Passive Optical Network\\
\textbf{GPR}	&Gaussian processes nonlinear regression\\
\textbf{HMM}	&Hidden Markov Model\\
\textbf{IP}		&Internet Protocol\\
\textbf{ISI}	&Inter-Symbol Interference\\
\textbf{LDE}	&Laser drift estimator\\
\textbf{MABP}	&Multi-arm bandit problem\\
\textbf{MDP}	&Markov decision processes\\
\textbf{MF}		&Modulation Format\\
\textbf{MFR}	&Modulation Format Recognition\\
\textbf{ML}		&Machine Learning\\
\textbf{MLP}	&Multi-layer perceptron\\
\textbf{MPLS}	&Multi-Protocol Label Switching\\
\textbf{MTTR}	&Mean Time To Repair\\
\textbf{NF}		&Noise Figure\\
\textbf{NFDM}	&Nonlinear Frequency Division Multiplexing\\
\textbf{NFT}	&Nonlinear Fourier Transform\\
\textbf{NLI}	&Nonlinear Interference\\
\textbf{NMF}	&Non-negative Matrix Factorization\\
\textbf{NN}		&Neural Network\\
\textbf{NPDM}	&Network planner and decision maker\\
\textbf{NRZ}	&Non-Return to Zero\\
\textbf{NWDM}	&Nyquist Wavelength Division Multiplexing\\
\textbf{OBS}	&Optical Burst Switching\\
\textbf{ODB}	&Optical Dual Binary\\
\textbf{OFDM}	&Orthogonal Frequency Division Multiplexing\\
\textbf{ONT}	&Optical Network Terminal\\
\textbf{ONU}	&Optical Network Unit\\
\textbf{OOK}	&On-Off Keying\\
\textbf{OPM}	&Optical Performance Monitoring\\
\textbf{OSNR}	&Optical Signal-to-Noise Ratio\\
\textbf{PAM}	&Pulse Amplitude Modulation\\
\textbf{PDL}	&Polarization-Dependent Loss\\
\textbf{PM} 	&Polarization-multiplexed\\
\textbf{PMD} 	&Polarization Mode Dispersion\\
\textbf{POI}	&Point of Interest\\
\textbf{PON}	&Passive Optical Network\\
\textbf{PSK}	&Phase Shift Keying\\
\textbf{QAM}	&Quadrature Amplitude Modulation\\
\textbf{Q-factor}	&Quality factor\\
\textbf{QoS}	&Quality of Service\\
\textbf{QoT}	&Quality of Transmission\\
\textbf{QPSK}	&Quadrature Phase Shift Keying\\
\textbf{RF}		&Random Forest\\
\textbf{RL}		&Reinforcement Learning\\
\textbf{RNN}	&Recurrent Neural Network\\
\textbf{ROADM}	&Reconfigurable Optical Add/Drop Multiplexer\\
\textbf{ROC}	&Receiver operating characteristic\\
\textbf{RWA}	&Routing and Wavelength Assignment\\
\textbf{RZ}		&Return to Zero\\
\textbf{S-BVT} 	&Sliceable Bandwidth Variable Transponders\\
\textbf{SDN} 	&Software-defined Networking\\
\textbf{SDON} 	&Software-defined Optical Network\\
\textbf{SLA} 	&Service Level Agreement\\
\textbf{SNR}	&Signal-to-Noise Ratio\\
\textbf{SPM}	&Self-Phase Modulation\\
\textbf{SSC}	&Signal spectrum comparison\\
\textbf{SSFM}	&Split-Step Fourier Method\\
\textbf{SSV}	&Signal spectrum verification\\
\textbf{SVM}	&Support Vector Machine\\
\textbf{TCP}	&Transmission Control Protocol\\
\textbf{TN}		&True negatives\\
\textbf{TP}		&True positives\\
\textbf{VNT}	&Virtual Network Topology\\
\textbf{VT}		&Virtual Topology\\
\textbf{VTD}	&Virtual Topology Design\\
\textbf{WDM}	&Wavelength Division Multiplexing\\
\textbf{XPM}	&Cross-Phase Modulation\\
\end{supertabular}